\documentclass{jfm}

\usepackage{graphicx}
\usepackage{newtxtext}
\usepackage{newtxmath}
\usepackage{natbib}
\usepackage{hyperref}
\usepackage{longtable}
\usepackage{ragged2e} 
\usepackage{booktabs,makecell, multirow, tabularx}
\hypersetup{
    colorlinks = true,
    urlcolor   = blue,
    citecolor  = black,
}

\newcommand{\RomanNumeralCaps}[1]
\linenumbers

\usepackage{xcolor}
\usepackage{amsmath}

\title{Temporal modulation on mixed convection in turbulent channels}

\author{Ao Xu\aff{1}$^,$\aff{2}$^,$\aff{3},
  Rui-Qi Li\aff{1}
 \and Heng-Dong Xi\aff{1}$^,$\aff{2}$^,$\aff{3}\corresp{\email{hengdongxi@nwpu.edu.cn}}}

\affiliation{\aff{1}Institute of Extreme Mechanics, School of Aeronautics, Northwestern Polytechnical University, Xi'an 710072, China
\aff{2}National Key Laboratory of Aircraft Configuration Design, Xi'an 710072, China
\aff{3}Key Laboratory for Extreme Mechanics of Aircraft of Ministry of Industry and Information Technology, Xi'an 710072, China
}

\begin{document}
\maketitle

\begin{abstract}
We studied flow organization and heat transfer properties in mixed turbulent convection within Poiseuille-Rayleigh-B\'enard channels subjected to temporally modulated sinusoidal wall temperatures. 
Three-dimensional direct numerical simulations were performed for Rayleigh numbers in the range $10^6  \leq Ra \leq 10^8$, a Prandtl number $Pr = 0.71$ and a bulk Reynolds number $Re_b \approx 5623$. 
We found that high-frequency wall temperature oscillations had minimal impact on flow structures, while low-frequency oscillations induced adaptive changes, forming stable stratified layers during cooling. 
Proper orthogonal decomposition (POD) analysis revealed a dominant streamwise unidirectional shear flow mode. 
Large-scale rolls oriented in the streamwise direction appeared as higher POD modes and were significantly influenced by lower-frequency wall temperature variations. 
Long-time-averaged statistics showed that the Nusselt number increased with decreasing frequency by up to 96\%, while the friction coefficient varied by less than 15\%. 
High-frequency modulation predominantly influenced near-wall regions, enhancing convective effects, whereas low frequencies reduced these effects via stable stratified layer formation. 
Phase-averaged statistics showed that high-frequency modulation resulted in phase-stable streamwise velocity and temperature profiles, while low-frequency modulation caused significant variations due to weakened turbulence. 
Turbulent kinetic energy (TKE) profiles remained high near the wall during both heating and cooling at high frequency, but decreased during cooling at low frequencies. 
A TKE budget analysis revealed that during heating, TKE production was dominated by shear near the wall and by buoyancy in the bulk region; 
while during cooling, the production, distribution and dissipation of TKE were all nearly zero. 
\footnote{
This article may be downloaded for personal use only.
Any other use requires prior permission of the author and Cambridge University Press.
This article appeared in Xu \emph{et al.}, J. Fluid Mech. \textbf{1006}, A11 (2025) and may be found at \url{https://doi.org/10.1017/jfm.2025.22}.
}
\end{abstract}


\section{Introduction}
\label{sec:1 Introduction}
Mixed convection, driven by both shear and buoyancy forces, occurs ubiquitously in nature and has widespread applications in industry \citep{caulfield2021layering}. 
For example, this complex interaction is essential for understanding the behaviour of atmospheric currents, where stratification can be stable (warmer air above cooler layers), unstable (lower layers are heated and rising) or neutral (temperature gradient minimal affecting air movements) \citep{zhang2023multifield,zhang2024structure}. 
During the day, unstable stratification can form large longitudinally aligned rollers \citep{brown1980longitudinal,young2002rolls,dror2023convective}. 
These structures generate rows of cumulus clouds and create striped patterns on sand dunes \citep{hanna1969formation,andreotti2009giant,kok2012physics}. 
At night, the atmospheric boundary layer is usually stably stratified, with temperature increasing with height, inhibiting vertical mixing and resulting in a more layered, less turbulent atmosphere \citep{nieuwstadt1984turbulent}. 
Another example is in nuclear engineering, where mixed convection is crucial for designing and operating Generation IV nuclear reactors, such as sodium-cooled and lead-cooled fast reactors. 
A key component of these reactors is the primary circuit, where heat generated from nuclear fission is transferred to a coolant before being converted to steam for electricity generation \citep{komen2023status}. 
Paradigms for studying mixed convection include the Poiseuille-Rayleigh-B\'enard (PRB) and the Couette-Rayleigh-B\'enard (CRB) systems, which combine Poiseuille (or Couette) flow with Rayleigh-B\'enard (RB) convective flow. 
Although extensive efforts have been devoted to studying shear-driven wall turbulence in Poiseuille flow (or Couette flow) systems \citep{marusic2010wall,smits2011high,jimenez2012cascades,jimenez2013near,graham2021exact,marusic2021energy,chen2022reynolds,yao2022direct}, and buoyancy-driven thermal turbulence in RB systems \citep{ahlers2009heat,lohse2010small,chilla2012territoriality,wang2020vibration,jiang2020supergravitational,xia2023tuning,lohse2024ultimate,lohse2024asking}, the interplay between horizontal shear and vertical buoyancy in mixed convection remains relatively less understood.

In turbulent channel flows with stable temperature stratification, fluid density increases with depth, and buoyancy forces act to return displaced fluid parcels to their original position, causing oscillations around the equilibrium point and forming internal gravity waves \citep{zonta2022interaction}. 
In contrast, in turbulent channel flows with unstable temperature stratification, thermal plumes significantly influence momentum and heat transport from the wall \citep{komori1982turbulence,iida1997direct}. 
When heat conduction in the bottom wall is coupled with fluid flow in the channel, the thermal properties and thickness of the conducting solid wall strongly affect the solid-fluid interfacial temperatures \citep{garai2014flow}. 
\citet{pirozzoli2017mixed} and \citet{blass2020flow,blass2021effect} found that at high friction Reynolds number ($Re_{\tau}$, describing the shear strength) and high Rayleigh number ($Ra$, describing the buoyancy strength) in both PRB and CRB systems, large-scale quasistreamwise roll structures form, occupying the entire channel height, a behaviour not observed in pure turbulent channel flow or pure turbulent RB convection. 
Using the direct numerical simulation (DNS) data of \citet{pirozzoli2017mixed}, \citet{madhusudanan2022navier} developed a linearized Naver-Stokes-based model that accurately captures the trends of these quasistreamwise rolls, emphasizing the significant impact of the relative effect of shear and buoyancy (characterized by the Richardson number $Ri_b$) on the predicted coherent structures. 
Meanwhile, \citet{cossu2022onset} showed that the linear instability of turbulent mean flow to coherent perturbations is linked to the onset of large-scale rolls and predicted the critical Rayleigh number for their formation.
Recently, \citet{howland2024turbulent} studied  a differentially heated vertical channel subject to a Poiseuille-like horizontal pressure gradient, which is relevant to industrial heat exchangers in applied thermal engineering.

In the exploration of turbulent flows driven by time-dependent forcing, such as atmospheric circulation driven by the Sun’s radiation causing daily warming and cooling cycles, efforts have been made to investigate temporally modulated turbulent RB convection. 
For example, \citet{jin2008experimental} experimentally imposed periodic pulses of energy to drive the convective flow, achieving a 7\% increase in heat transfer efficiency as measured by the Nusselt number ($Nu$). 
\citet{niemela2008formation} experimentally assessed the flow structure under sinusoidal heating modulation at the lower boundary and discovered a core region with near-superconducting behaviour, where thermal waves propagate without attenuation. 
Due to experimental challenges in achieving a broad range of modulation frequencies, \citet{yang2020periodically} performed DNS over four orders of magnitude of modulation frequencies and reported an appreciable enhancement of $Nu$ by up to 25\%. 
Using the concept of the Stokes thermal boundary layer, they explained the onset frequency of the $Nu$ enhancement and the optimal frequency at which $Nu$ is maximal. 
Later, \citet{urban2022thermal} used helium gas in cryogenic conditions to create convection cells that respond quickly to temperature changes, allowing them to achieve high modulation amplitudes and a wide range of frequencies. 
Their results confirmed  the numerical predictions of \citet{yang2020periodically} regarding the significant enhancement of $Nu$. 
These advances in understanding temporally modulated buoyancy-driven natural convection lead to the question of how temporal modulation affects flow organization and heat transfer efficiency in mixed convection.

In this work, we aim to investigate the effects of temporal modulation on mixed convection in turbulent channels. 
Motivated by the studies of atmospheric currents in desert regions, we consider stable, unstable or neutral stratification conditions, achieved by temporally modulating the bottom wall temperature. 
Based on data from moderate-resolution imaging spectroradiometer observations \citep{sharifnezhadazizi2019global}, we applied sinusoidal modulation to the bottom wall to mimic diurnal variations in land surface temperature. 
By analysing turbulent quantities over time within the modulation period, we can unravel the transient mechanisms and phase dynamics of the flow structure \citep{manna2015pulsating,ebadi2019mean}. 
The rest of this paper is organized as follows. 
In $\S$ \ref{sec:2 Numerical}, we present numerical details for the DNS of mixed convection. 
In $\S$ \ref{sec:3 Results}, we first describe the instantaneous flow and heat transfer features in the temporally modulated mixed convection channel, followed by an analysis of long-time-averaged statistics and phase-averaged statistics. 
The long-time-averaged quantity is calculated as $\overline{\mathbb{F}}(\mathbf{x})=\lim_{N_{\text{total}}\to\infty}[\sum\mathbb{F}(\mathbf{x},t)]/N_{\text{total}}$, where $\mathbb{F}(\mathbf{x},t)$ is the instantaneous field and $N_{\text{total}}$ is the total snapshot number of the instantaneous field. 
The phase-averaged quantity $\left\langle\mathbb{F}\left(\mathbf{x}\right)\right\rangle_{\phi}$ is calculated as $\left\langle\mathbb{F}(\mathbf{x})\right\rangle_{\phi}=\lim_{N_{\text{cycles}}\to\infty}[\sum\mathbb{F}(\mathbf{x},t_k)]/N_{\text{cycles}}$, where $N_{\text{cycles}}$ is the number of cycles and $t_k$ are the times corresponding to the phase $\phi$.  
In $\S$ \ref{sec:4 Conclusion}, the main findings of the present work are summarized.

\section{Numerical method}
\label{sec:2 Numerical}
\subsection {Direct numerical simulation of thermal turbulence}
In incompressible thermal convection, we employ the Boussinesq approximation to account for temperature as an active scalar that influences the velocity field through buoyancy effects in the vertical direction, assuming constant transport coefficients. 
The flow is also driven by a body force (or equivalently a mean pressure gradient) that accounts for shear effects in the horizontal direction. 
The equations governing fluid flow and heat transfer can be written as
\begin{equation}\label{massequation}
	\nabla  \cdot {{\boldsymbol{u}}} = 0
\end{equation}
\begin{equation}\label{momentumequation}
\frac{\partial\mathbf{u}}{\partial t}+\mathbf{u}\cdot\nabla\mathbf{u}=-\frac1{\rho_0}\nabla P+\nu\nabla^2\mathbf{u}+f_b\hat{\mathbf{x}}+g\beta(T-T_0)\hat{\mathbf{y}}
\end{equation}
\begin{equation}\label{energyequation}
	\frac{{\partial T}}{{\partial t}} + {{\boldsymbol{u}}} \cdot \nabla T = \alpha {\nabla ^2}T
\end{equation}
Here, ${{\boldsymbol{u}}}$ is the fluid velocity and  $P$ and ${T}$  are the pressure and temperature of the fluid, respectively. 
The coefficients $\beta$, $\nu$ and $\alpha$  denote the thermal expansion coefficient, kinematic viscosity and thermal diffusivity, respectively. 
Reference state variables are indicated by subscripts zeros.  
The vectors $\hat{\mathbf{x}}$ and $\hat{\mathbf{y}}$ are unit vectors in the streamwise and wall-normal directions, respectively.
The term $g$ represents the gravitational acceleration in the wall-normal direction. 
The term $f_b$ represents a body force used to maintain a constant bulk flow rate in the streamwise direction. 
This forcing is spatially uniform but time-dependent, allowing precise control over the flow rate at each time step \citep{pirozzoli2017mixed}. 
While a constant pressure gradient is commonly used to drive flow in channel simulations, in the presence of buoyancy forces, a constant pressure gradient can lead to variations in the bulk flow rate, as buoyancy may either enhance or oppose the mean flow depending on the temperature distribution. 
By maintaining a constant flow rate in mixed convection, we can use the mean flow strength as a control parameter, allowing us to effectively attribute changes in flow behaviour to specific factors such as thermal stratification or flow strength.

We introduce the non-dimensional variables
\begin{equation}
	\begin{split}
		&{{\boldsymbol{x}}^*} = {\boldsymbol{x}}/H,\;\;\;{t^*} = t/ ({H/u_b}) ,\;\;\;{\boldsymbol{u}}^* = {{\boldsymbol{u}}}/u_b \\
		{P^*} = P/&({\rho _0}{u_b}^2),\;\;\;T^* = ({T} - {T_0})/{\Delta _T},\;\;\;{f_b}^* = f_b/({{u_b}^2/H})
	\end{split}
\end{equation}
where $u_b$ is the bulk velocity, $H$ denotes the channel height and $\Delta_T = T_{\text{hot}}-T_{\text{cold}}$ is the temperature difference between the heating and cooling walls. 
We can rewrite (\ref{massequation})-(\ref{energyequation}) in dimensionless form:
\begin{equation}
	\nabla  \cdot {\boldsymbol{u}}^* = 0
\end{equation}
\begin{equation}
\frac{\partial\mathbf{u}^*}{\partial t^*}+\mathbf{u}^*\cdot\nabla\mathbf{u}^*=-\nabla P^*+\frac1{Re_b}\nabla^2\mathbf{u}^*+f_b^*\hat{\mathbf{x}}+\frac{Ra}{Re_b^2Pr}T^*\hat{\mathbf{y}}
\end{equation}
\begin{equation}
	\frac{{\partial {T^*}}}{{\partial {t^*}}} + {{\boldsymbol{u}}^*} \cdot \nabla {T^*} =  \frac{1}{{Re_b Pr}} {\nabla ^2}{T^*}
\end{equation}
Here, the control dimensionless parameters include the bulk Reynolds number ($Re_{b}$), Rayleigh number ($Ra$) and Prandtl number ($Pr$), which are defined as
\begin{equation}
	Re_b = \frac{Hu_b}{\nu},\;\;\;Ra = \frac{{g\beta {\Delta _T}{H^3}}}{{\nu \alpha }},\;\;\;Pr = \frac{\nu }{\alpha }
\end{equation}

The global competition between shear and buoyancy effects can be quantified by the bulk Richardson number ($Ri_b$) as $Ri_b=Ra/(Re_b^2 Pr)$.
The extreme case of $Ri_b = 0$ represents purely shear driving and $Ri_b=\infty$ represents purely buoyancy driving.
We adopt the finite volume method (FVM) implemented in the open-source OpenFOAM solver (Version 8) for DNS. 
Specifically, we use the transient \emph{buoyantPimpleFoam} solver with the turbulence model turned off. 
Convective terms and viscous terms are discretized using a second-order central differencing scheme, while temporal terms are discretized using a second-order implicit backward differencing scheme based on three-time levels. 
Pressure-velocity coupling is achieved with the pressure-implicit splitting of operators (PISO) algorithm, with PISO corrections set to four, following the settings by \citet{komen2014quasi}. 
For the momentum equation, we use a preconditioned biconjugate gradient method designed for asymmetric matrices, along with diagonal-based incomplete LU preconditioning. 
The pressure equation is solved using the geometric agglomerated algebraic multigrid method. 
Time advancement is controlled by adaptive time stepping, with the adaptive time step regulated by the cell-face Courant number, keeping the maximum cell-face Courant numbers below 0.5. 
More numerical details and validation of the OpenFOAM solver can be found in \citet{komen2014quasi,komen2017quantification} and \citet{kooij2018comparison}. 
To verify our OpenFOAM results, we also conducted simulations at $Re_b \approx 3162$, $Ra = 10^7$ and $Pr = 1$ using an in-house solver based on the lattice Boltzmann method (LBM) \citep{xu2017accelerated,xu2019lattice,xu2023multi}. 
The results from both the open-source OpenFOAM solver and the in-house lattice Boltzmann solver were consistent, as discussed in appendix \ref{appA}.

\subsection {Simulation settings}

\begin{figure}
	\centerline{\includegraphics[width=0.45\textwidth]{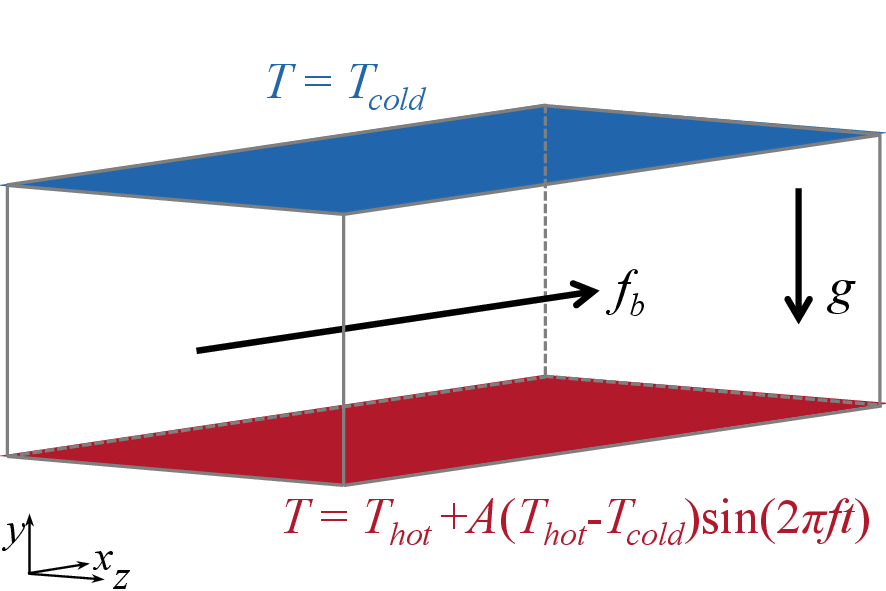}}
	\caption{Schematic illustration of the temporally modulated mix convection channel.}
	\label{fig:demo}
\end{figure}

We explore the dynamics of mixed convection within a three-dimensional (3-D) channel of dimensions $L \times H \times W$, as illustrated in figure \ref{fig:demo}. 
Here, $L$ is the length, $H$ is the height and $W$ is the width of the simulation domain, 
$x$ denotes the streamwise direction, $y$ denotes the wall-normal direction and $z$ denotes the spanwise direction. 
The computational domain size is chosen as $2\pi h \times 2h \times \pi h$ to ensure the spanwise domain size is close to the minimal spanwise size required to achieve developed turbulence in mixed convection simulations \citep{pirozzoli2017mixed}. 
Here, $h=H/2$ is the half-height of the channel.
Periodic boundary conditions for velocity and temperature are applied in the streamwise and spanwise directions to exploit statistical homogeneity. 
In the wall-normal direction, no-slip velocity boundary conditions are imposed on the top and bottom walls of the channel. 
The grid size distribution is symmetric about the mid-plane, with cell sizes increasing geometrically from the wall to the bulk using a geometric progression.
Specifically, the cell size for the $i$th grid $\Delta y_{i}$ can be defined as $\Delta y_{i} = \Delta y_{\text{wall}} q^{i-1}$ (for $i=1,2,\cdots,M$), where $\Delta y_{\text{wall}}$ is the starting cell size, $q$ is the common ratio of the geometric progression and $M$ is the number of cells in the half-channel.
The expansion ratio $r$ of the cell size, defined as the ratio of the end cell size $\Delta y_{\text{bulk}}$ to the starting cell size $\Delta y_{\text{wall}}$, is expressed as $r=\Delta y_{\text{bulk}} / \Delta y_{\text{wall}}$.
This gives $q=r^{1/(M-1)}$ and $\Delta y_{\text{wall}}=h(1-q)/(1-q^M)$.
We set $r=8$ for $Ra = 10^6$ and $r=5$ for $Ra = 10^7$ and $10^8$ in the simulations. 
We referenced the grid set-up used by \citet{pirozzoli2017mixed}, which is based on the resolution requirements for pure buoyant flow \citep{shishkina2010boundary} and pure channel flow \citep{bernardini2014velocity}. 
We performed an \emph{a posteriori} validation for all simulation cases, ensuring that the grid size in each coordinate direction is less than three Kolmogorov units in all cases.

\begin{table}
	\begin{center}
		\def~{\hphantom{0}}
		\begin{tabular}{cccccc}
			$Ra$	&  $Ri_b$ & $f^{*}$ & $Re_{\tau}$ (bottom/top) & $Re_{\tau}$ difference (\%) & $N_x \times N_y \times N_z$  \\
			$10^6$  &  0.0445 & Unmodulated  & $175.05 / 176.03$ & $0.56$ & $192 \times 156 \times 96$\\
			$10^6$	&  0.0445 & 1    & $175.44 / 175.75$ & $0.18$ & $192 \times 156 \times 96$\\
			$10^6$	&  0.0445 & 0.1  & $169.02 / 179.75$ & $6.35$ & $192 \times 156 \times 96$\\
			$10^6$	&  0.0445 & 0.01 & $167.82 / 178.10$ & $6.13$ & $192 \times 156 \times 96$\\
			$10^7$	&  0.445  & Unmodulated  & $197.90 / 198.71$ & $0.41$ & $384 \times 256 \times 192$\\
			$10^7$	&  0.445  & 1    & $197.50 / 197.35$ & $0.08$ & $384 \times 256 \times 192$\\
			$10^7$	&  0.445  & 0.1  & $202.33 / 212.50$ & $5.03$ & $384 \times 256 \times 192$\\
			$10^7$	&  0.445  & 0.01 & $187.11 / 214.35$ & $14.56$& $384 \times 256 \times 192$\\
			$10^8$	&  4.45   & Unmodulated  & $272.43 / 273.01$ & $0.21$ & $384 \times 256 \times 192$\\
			$10^8$	&  4.45   & 1    & $270.25 / 272.21$ & $0.73$ & $384 \times 256 \times 192$\\
			$10^8$	&  4.45   & 0.1  & $266.75 / 286.46$ & $7.39$ & $384 \times 256 \times 192$\\
			$10^8$	&  4.45   & 0.01 & $251.19 / 295.31$ & $15.56$& $384 \times 256 \times 192$\\
		\end{tabular}%
		\caption{Numerical details of flow quantities. The columns from left to right indicate the following: Rayleigh number $Ra$; bulk Richardson number $Ri_b$; wall-temperature modulation frequency $f^{*}$; friction Reynolds number $Re_{\tau}$ at the bottom and top wall, respectively, and their relative differences due to asymmetric boundary conditions; grid number $N_x \times N_y \times N_z$.}
		\label{tab:Numerical details}
	\end{center}
\end{table}

For the temperature boundary condition, the top wall is maintained at a constant low temperature of $T_{\text{top}}=T_{\text{cold}}$, while the bottom wall is subjected to a sinusoidal modulation $T_{\text{bottom}}=T_{\text{hot}}+A(T_{\text{hot}}-T_{\text{cold}})\sin(2\pi ft)$. 
Here, $A$ is the modulation amplitude, $f$ is the modulation frequency and $t$ is time.
We define the phase angle of the modulation as $\phi=2\pi ft$, and this phase angle $\phi$ will be used to describe the temporal progression of the modulation.
The modulation amplitude is fixed as $A = 2$, resulting in a temperature difference $\Delta_T^*=\left(T_{\text{bottom}}-T_{\text{top}}\right)/\left(T_{\text{hot}}-T_{\text{cold}}\right)=1+2\sin\left(2\pi ft\right) $, covering the regimes of unstable stratification ($\Delta_{T} > 0$), neutral stratification ($\Delta_{T}=0$) and stable stratification ($\Delta_{T}<0$). 
Previously, \citet{yang2020periodically} fixed the modulation amplitude as $A = 1$ in RB convection, which did not explore the stable stratification cases because $A = 1$ leads to $\Delta_{T}\geq 0$ . 
We adopt the buoyancy timescale $t_{f}=\sqrt{H/(g\beta \Delta_{T})}$ to non-dimensionalize the modulation period $T_{\text{period}}$ as $T_{\text{period}}^{*}=T_{\text{period}}/t_{f}$ \citep{yang2020periodically}.
We set the dimensionless modulation frequency $f^{*}$ in the range $0.01\leq f^{*}=1/T_{\text{period}}^{*}=t_{f}/T_{\text{period}} \leq 1$. 
Another approach is to adopt the bulk convective timescale $t_{b}=H/u_{b}$ to non-dimensionalize the modulation period as $T_{\text{period}}^{*}=T_{\text{period}}/t_{b}$.
In this case, the modulation frequency is given by $f^{*}_{b}=t_{b}/T_{\text{period}}=f^{*}t_{b}/t_{f}$, leading to $f^{*}_{b}=\sqrt{Ri_{b}}f^{*}$.
Because the wall temperature modulation changes the buoyancy within the flow system, we will focus on discussing $f^{*}$ in the following analysis.

To determine the Rayleigh number, the time-averaged temperature difference $\langle{\Delta_T} \rangle_{t}=T_{\text{hot}}-T_{\text{cold}}$ is adopted as $Ra=g\beta\langle{\Delta_T} \rangle_{t} H^3/(\nu\alpha)$, and the Rayleigh number is in the range of $10^6 \leq Ra \leq 10^8$. 
We fix the Prandtl number as $Pr = 0.71$ for the working fluid of air. 
The bulk Reynolds number is $Re_b =10^{3.75} \approx  5623$, and the corresponding friction Reynolds number $Re_{\tau}=u_{\tau}h/\nu$ is also provided in table \ref{tab:Numerical details}, where $u_{\tau}=\sqrt{\langle{\tau_{w}}\rangle /\rho}$ is the friction velocity associated with the mean wall shear stress $\langle{\tau_{w}}\rangle$.
We are particularly interested in the mixed convection regime, such that the Richardson number is in the range of $0.0445 \leq Ri_b \leq 4.45$. 
A list of the bulk flow parameters obtained from all the simulations conducted in this study is provided in table \ref{tab:Numerical details}, and their relevance to atmospheric currents in desert regions is discussed in appendix \ref{appB}.

\section{Results and discussion}
\label{sec:3 Results}
\subsection {Instantaneous flow and heat transfer features}

\begin{figure}
	\centerline{\includegraphics[width=0.8\textwidth]{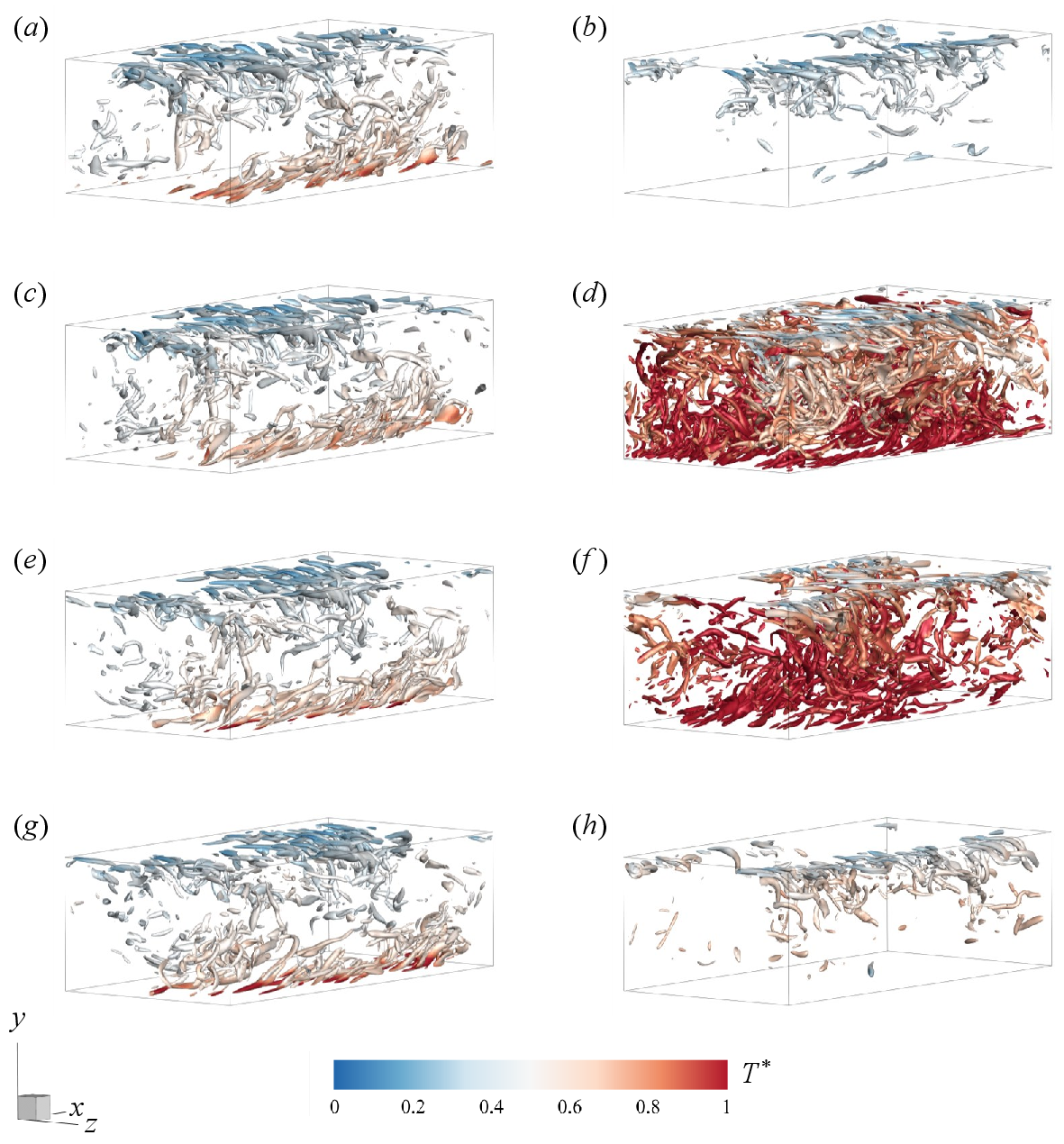}}
	\caption{Typical instantaneous isosurfaces of the $Q$-criterion, $Q^* = \left(\|\mathbf{\Omega}^*\|^2-\|\mathbf{S}^*\|^2 \right)/2 = 15$ coloured by the local temperature $T^*$, 
at phase angle (\emph{a},\emph{b}) $\phi = 0$, (\emph{c},\emph{d}) $\phi = \pi/2$, (\emph{e},\emph{f}) $\phi = \pi$, (\emph{g},\emph{h}) $\phi = 3\pi/2$, 
with frequency (\emph{a},\emph{c},\emph{e},\emph{g}) $f^* = 1$, (\emph{b},\emph{d},\emph{f},\emph{h}) $f^* = 0.01$, for $Ra = 10^7$ and $Re_b \approx 5623$.}
	\label{fig:Q}
\end{figure}

In figure \ref{fig:Q}, we show snapshots of isosurfaces of the $Q$-criterion, coloured by the local temperature $T^*$, and the corresponding video can be viewed in supplementary movie 1 available at \url{https://doi.org/10.1017/jfm.2025.22}.
The $Q$-criterion visualizes vortices where the magnitude of vorticity $\mathbf{\Omega}=[\nabla \mathbf{u}-(\nabla \mathbf{u})^T]/2$ exceeds the magnitude of the strain rate $\mathbf{S}=[\nabla \mathbf{u}+(\nabla \mathbf{u})^T]/2$, making it an effective tool for illustrating vortical structures \citep{hunt1988eddies}. 
At a higher modulation frequency of $f^* = 1$ (see figure \ref{fig:Q}\emph{a},\emph{c},\emph{e},\emph{g}), the vortex structures are concentrated near the wall and appear relatively unchanged across different phase angles. 
At a lower modulation frequency of $f^* = 0.01$ (see figure \ref{fig:Q}\emph{b},\emph{d},\emph{f},\emph{h}), the effect of temporal modulation on wall temperature is evident, and the vortex structure exhibits a pronounced temporal evolution. 
Specifically, an increased number of structures are identified at $\phi = \pi/2$ and $\phi = \pi$, carried by hot fluids. 
An interesting observation is that when the wall temperature modulation leads to a stable condition (see figure \ref{fig:Q}\emph{h}), the vortices near the bottom wall almost disappear. 
The absence of vortex interaction implies that the turbulent flow near the bottom wall is significantly weakened, although vortices still emerge and turbulent flow remains active on the top wall. 
The slower modulation frequency allows for the development of diverse flow structures and temperature distributions at each phase, indicating the fluid's ability to adapt to each state of the heating and cooling cycles. 
In contrast, rapid modulation does not give the fluid enough time to significantly rearrange its structure within one modulation cycle, resulting in a consistent pattern regardless of the phase angle.

\begin{figure}
	\centerline{\includegraphics[width=0.8\textwidth]{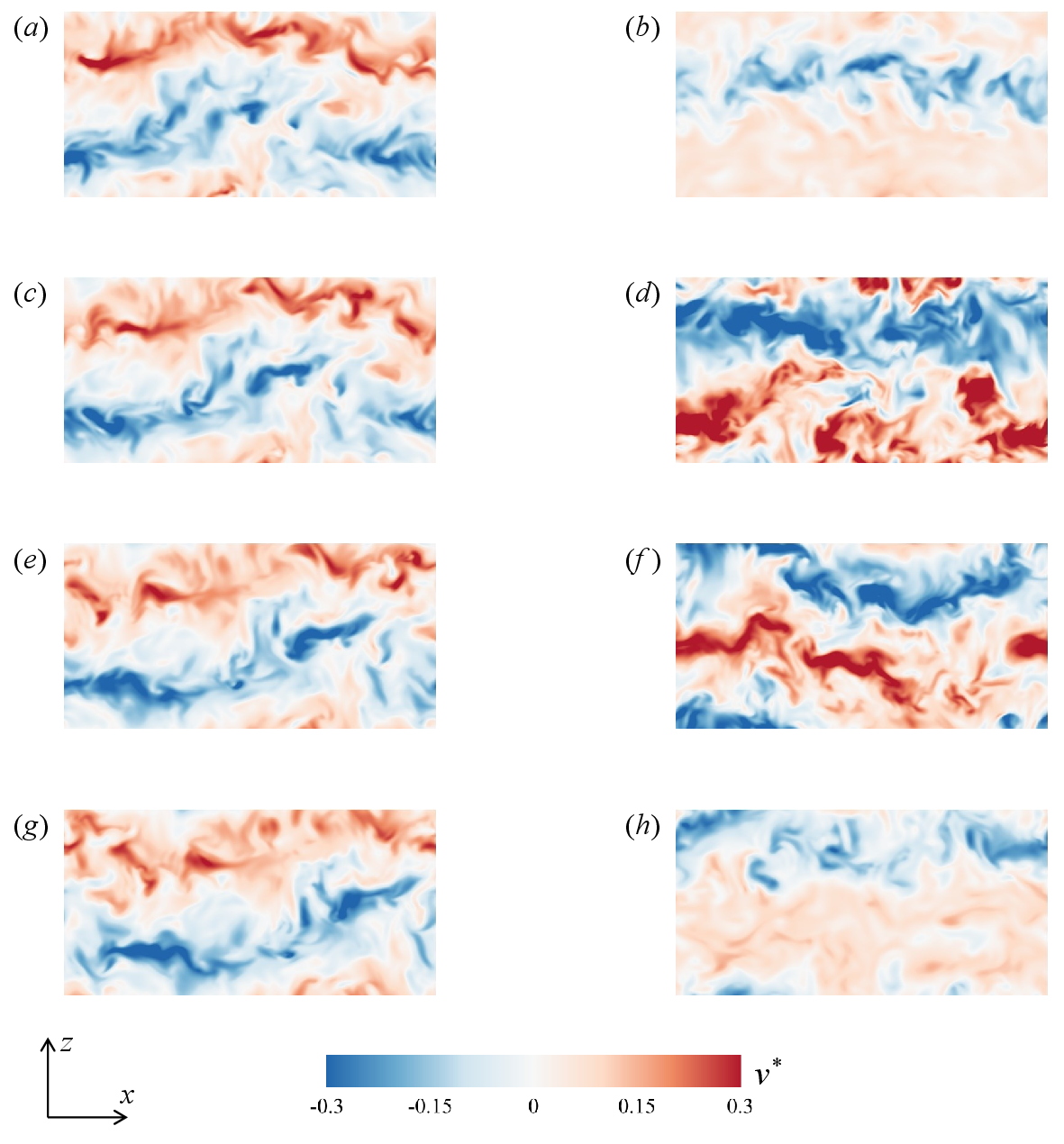}}
	\caption{Typical instantaneous velocity component $v^*$ at channel centre plane ($y = h$) at phase angle (\emph{a},\emph{b}) $\phi = 0$, (\emph{c},\emph{d}) $\phi = \pi/2$, (\emph{e},\emph{f}) $\phi = \pi$, (\emph{g},\emph{h}) $\phi = 3\pi/2$, 
with frequency (\emph{a},\emph{c},\emph{e},\emph{g}) $f^* = 1$, (\emph{b},\emph{d},\emph{f},\emph{h}) $f^* = 0.01$, for $Ra = 10^7$ and $Re_b \approx 5623$.}
	\label{fig:V_at_yh}
\end{figure}

We show the instantaneous velocity component $v^{*}$ at the channel centre plane ($y = h$) in figure \ref{fig:V_at_yh}, corresponding to the instantaneous state presented in figure \ref{fig:Q}. 
In convective flow, rising fluids are warmer, and falling fluids are colder; thus, we do not show the corresponding temperature field $T^*$ at this plane. 
However, we have verified that the correlation coefficient between $v^*$ and $T^*$ is larger than 0.47. 
At a Rayleigh number of $Ra = 10^7$ and a bulk Reynolds number of $Re_b \approx 5623$, with a fixed Prandtl number of $Pr = 0.71$, the corresponding Richardson number is $Ri_{b} = 0.45$. 
At this intermediate Richardson number, the shear-driven and buoyancy-driven turbulence production rates are nearly balanced. 
The flow exhibits rolls pointing in the streamwise direction, with a strong meandering behaviour due to the wavy instability of the rolls. 
This overall trend is similar to that reported by \citet{pirozzoli2017mixed}. 
In addition, we found that at a higher modulation frequency of $f^* = 1$ (see figure \ref{fig:V_at_yh}\emph{a},\emph{c},\emph{e},\emph{g}), the pair of counter-rotating rolls within the flow domain remains relatively stable in strength. 
The stability of these rolls demonstrates the flow's resilience against high-frequency thermal perturbations, suggesting an inherent inertia in the thermal field that resists rapid changes. 
However, at a lower modulation frequency of $f^* = 0.01$ (see figure \ref{fig:V_at_yh}\emph{b},\emph{d},\emph{f},\emph{h}), the strength of the roll varies with the phase angle. 
Specifically, the rolls are stronger during the heating phase  ($T_{\text{bottom}}>T_{\text{hot}}$); they are much weaker, or even disappear, during the cooling phase ($T_{\text{bottom}}<T_{\text{hot}}$). 

\begin{figure}
	\centerline{\includegraphics[width=0.8\textwidth]{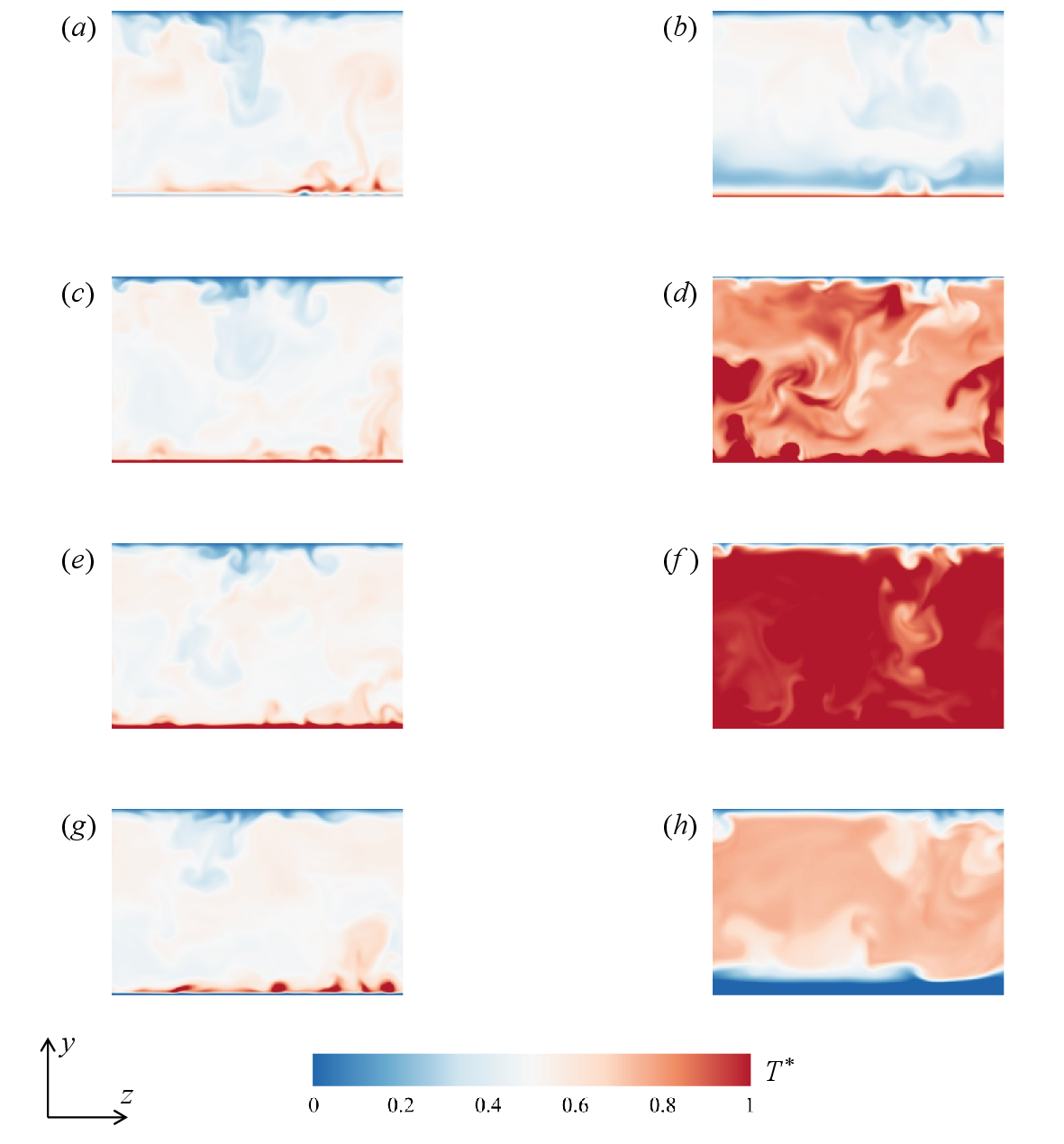}}
	\caption{Typical instantaneous temperature $T^*$ in cross-stream plane at phase angle (\emph{a},\emph{b}) $\phi = 0$, (\emph{c},\emph{d}) $\phi = \pi/2$, (\emph{e},\emph{f}) $\phi = \pi$, (\emph{g},\emph{h}) $\phi = 3\pi/2$, 
with frequency (\emph{a},\emph{c},\emph{e},\emph{g}) $f^* = 1$, (\emph{b},\emph{d},\emph{f},\emph{h}) $f^* = 0.01$, for $Ra = 10^7$ and $Re_b \approx 5623$.}
	\label{fig:T_at_x05Lx}
\end{figure}

We show the temperature field $T^*$ in the cross-stream plane, which complements the velocity contours at the channel centre plane shown in figure \ref{fig:V_at_yh}, and the corresponding video can be viewed in supplementary movie 2. 
During the heating phase, we observe frequent hot plume emissions near the bottom wall, becoming more pronounced after the wall temperature cycle reaches its peak (see figure \ref{fig:T_at_x05Lx}\emph{e},\emph{f}). 
In contrast, during the cooling phase (see figure \ref{fig:T_at_x05Lx}\emph{g},\emph{h}), plume emissions from the bottom wall are absent due to stable stratification, resulting in the weakening of buoyancy forces. 
In this stable stratification, internal gravity waves separate the bottom cold part and the top hot part \citep{zonta2022interaction}. 
We also observe the effect of modulation frequency on temperature evolution. 
At a higher frequency of $f^* = 1$, the upward- and downward-travelling plumes detaching from the boundary layers are almost unaffected by the phase angle, explaining the relatively stable pair of counter-rotating rolls found in figure \ref{fig:V_at_yh}(\emph{a},\emph{c},\emph{e},\emph{g}). 
At a lower frequency of $f^* = 0.01$, the slower frequency allows the temperature to respond to changes in the thermal boundaries and adapt to each state of the modulation cycle (see figure \ref{fig:T_at_x05Lx}\emph{b},\emph{d},\emph{f},\emph{h}). 
During the heating phase, hot rising plumes deeply penetrate into the bulk region of the channel. 
During the cooling phase, a stably stratified layer forms near the bottom wall, acting as a thermal blanket. 
The overall behaviour mirrors the dynamics observed in atmospheric flow \citep{dupont2022influence}. 
After sunrise, the ground heats up, destabilizing atmospheric conditions and forming a mixed layer with a relatively uniform temperature profile due to turbulent mixing. 
After sunset, the ground cools down more rapidly than the air above it, forming a stable boundary layer close to the surface. 
This layer sits beneath the remnants of the daytime mixed layer, where turbulence gradually decays in the absence of additional production mechanisms. 

\begin{figure}
 	\centerline{\includegraphics[width=0.8\textwidth]{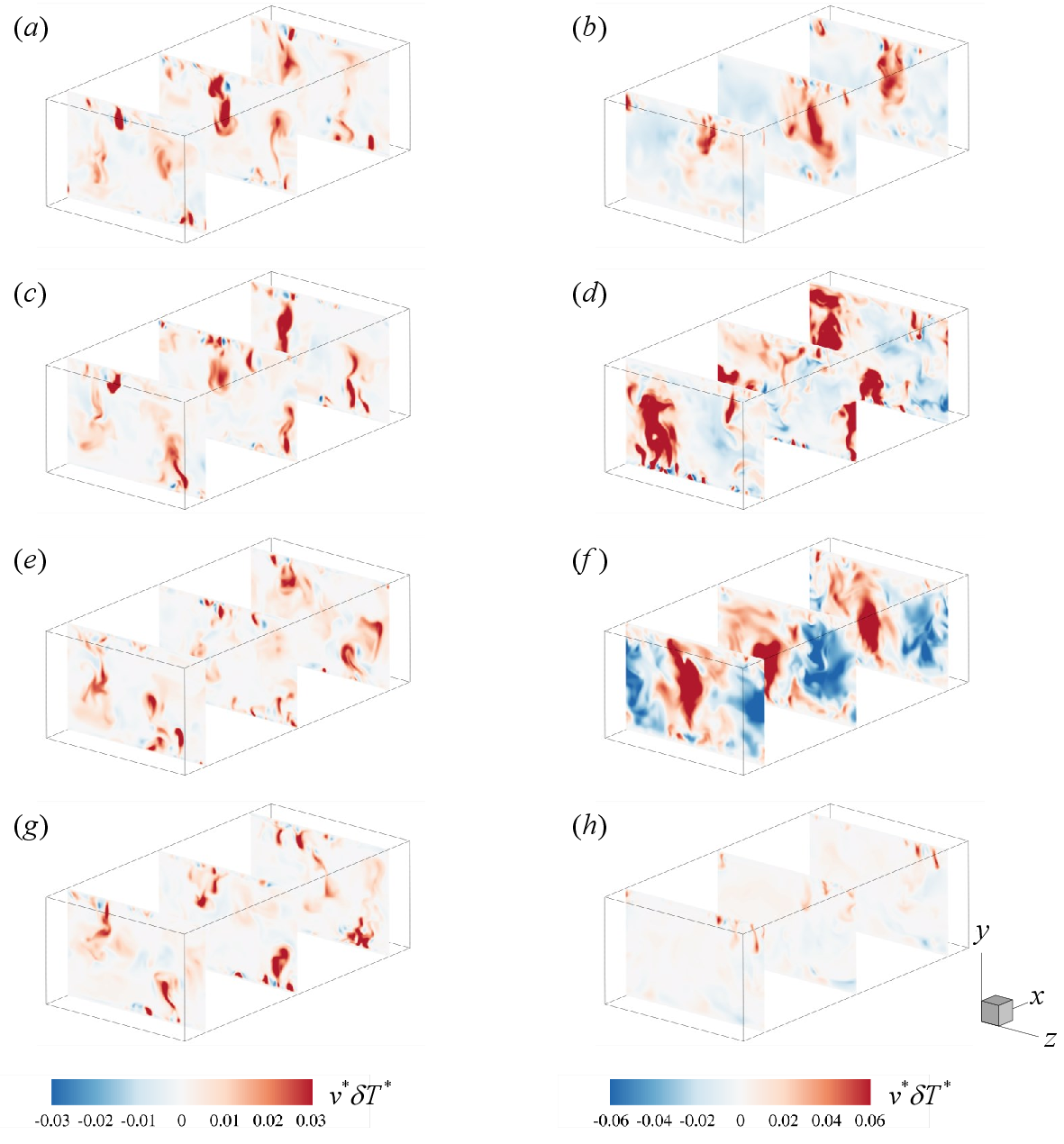}}
 	\caption{Typical instantaneous vertical convective heat flux $v^* \delta T^*$ at phase angle (\emph{a},\emph{b}) $\phi$ = 0, (\emph{c},\emph{d}) $\phi$ = $\pi$/2, (\emph{e},\emph{f}) $\phi$ = $\pi$, (\emph{g},\emph{h}) $\phi$ = 3$\pi$/2, 
 with frequency (\emph{a},\emph{c},\emph{e},\emph{g}) $f^* = 1$, (\emph{b},\emph{d},\emph{f},\emph{h}) $f^* = 0.01$, for $Ra = 10^7$ and $Re_b \approx 5623$.}
 	\label{fig:heatFluxSlice}
 \end{figure}

To examine the influence of wall temperature modulation on local heat transfer properties, we show slices of vertical convective heat flux $v^*\delta T^*$ in figure \ref{fig:heatFluxSlice}, corresponding to the instantaneous state presented in figure \ref{fig:Q}. 
Here, the temperature fluctuation is defined as $\delta T^* = T^* - \langle T \rangle_{V,t}$, and $\langle \cdots \rangle_{V,t}$ denotes the average over the whole channel and over the time.
At a higher frequency of $f^* = 1$ (see figure \ref{fig:heatFluxSlice}\emph{a},\emph{c},\emph{e},\emph{g}), positive values of vertical heat flux predominantly appear within the channel, while negative values occur near both the top and bottom walls. 
This counter-gradient heat transfer is attributed to sweeps of hotter fluid towards the bottom wall and colder fluid towards the top wall. 
In contrast, at a lower frequency of $f^* = 0.01$ (see figure \ref{fig:heatFluxSlice}\emph{b},\emph{d},\emph{f},\emph{h}), there is a strong counter-gradient heat flux in the bulk region of the channel, driven by the bulk dynamics of rolls, similar to the mechanism in RB convection \citep{gasteuil2007lagrangian,huang2013counter}. 
Detached plumes move with the streamwise-oriented roll, and after reaching the opposite wall, some plumes retain thermal energy and remain hotter or colder than their surroundings. 
They continue moving with the rolls, resulting in the falling of hot fluid or the rising of cold fluid, thus generating negative vertical heat flux.
 
\begin{figure}
	\centerline{\includegraphics[width=0.8\textwidth]{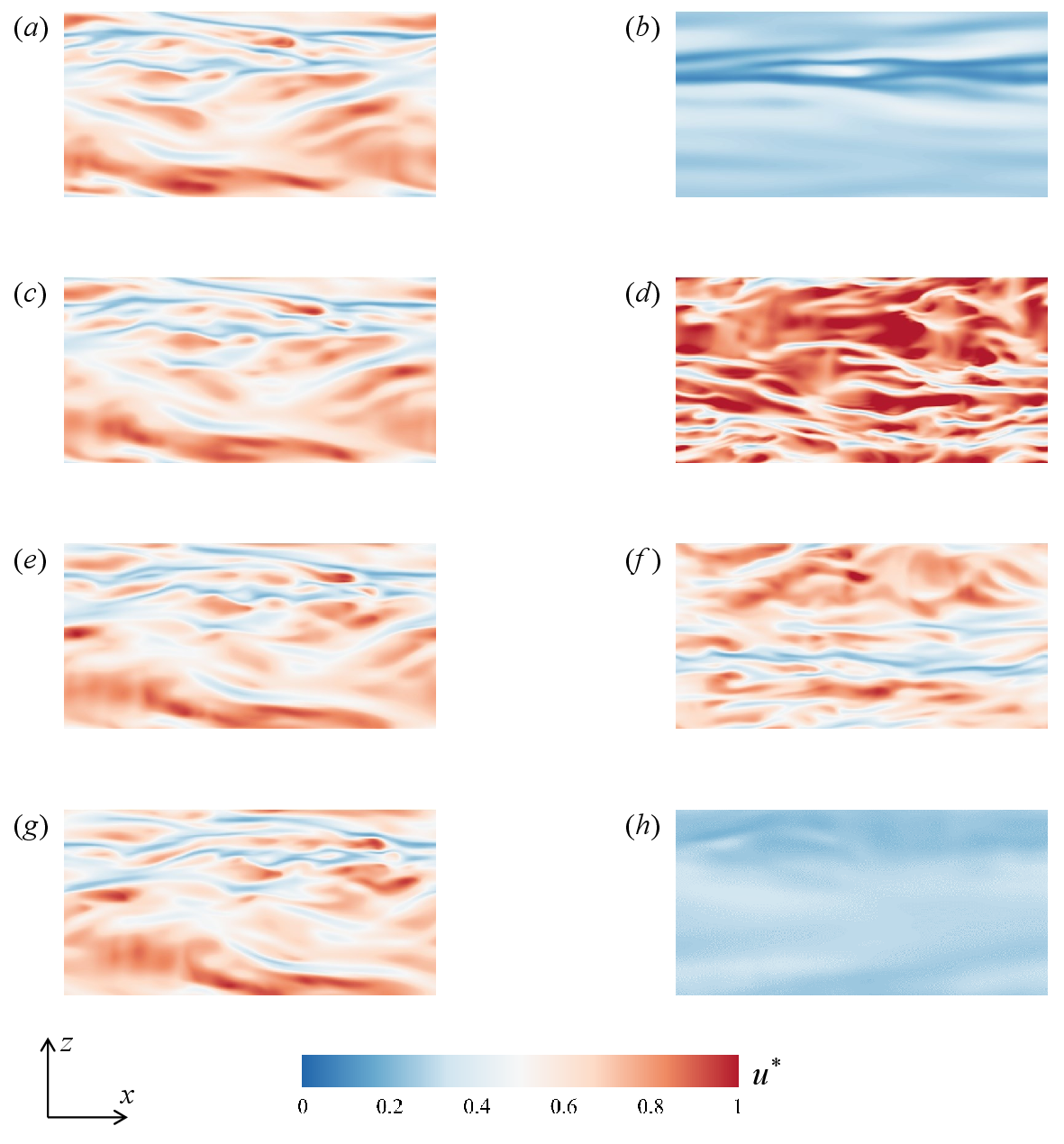}}
	\caption{Typical instantaneous velocity component $u^*$ at near-wall station ($y = 0.05 h$) at phase angle (\emph{a},\emph{b}) $\phi = 0$, (\emph{c},\emph{d}) $\phi = \pi/2$, (\emph{e},\emph{f}) $\phi = \pi$, (\emph{g},\emph{h}) $\phi = 3\pi/2$, 
with frequency (\emph{a},\emph{c},\emph{e},\emph{g}) $f^* = 1$, (\emph{b},\emph{d},\emph{f},\emph{h}) $f^* = 0.01$, for $Ra = 10^7$ and $Re_b \approx 5623$.}
	\label{fig:U_at_y005h}
\end{figure}

We further show the instantaneous streamwise velocity component $u^*$ at a near-wall station of $y = 0.05 h$ in figure \ref{fig:U_at_y005h}. 
At a higher modulation frequency of $f^* = 1$ (see figure \ref{fig:U_at_y005h}\emph{a},\emph{c},\emph{e},\emph{g}), near-wall streaks are observed even in the presence of strong buoyancy. 
These streaks are often associated with vigorous momentum transfer and robust turbulence production. 
However, at a lower modulation frequency of $f^* = 0.01$ (see figure \ref{fig:U_at_y005h}\emph{b},\emph{d},\emph{f},\emph{h}), during the cooling phase, which leads to stable stratification of the fluid, the near-wall burst-sweep process is completely disrupted (see figure \ref{fig:U_at_y005h}\emph{h}). 
This disruption ceases turbulence production and shows tendencies of relaminarization near the bottom wall. 
During the heating phase, which leads to unstable stratification, the near-wall streaks appear again. 

\begin{figure}
	\centerline{\includegraphics[width=0.95\textwidth]{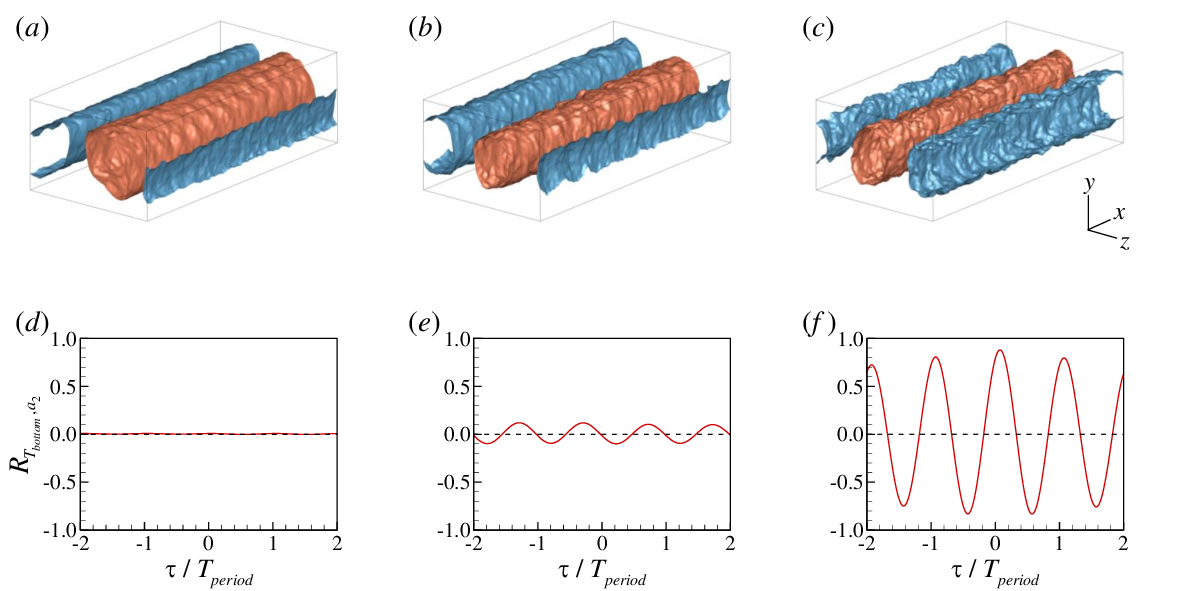}}
	\caption{(\emph{a}-\emph{c}) The second most energetic POD modes, visualizations of isosurfaces of vertical velocity $v$ (red colour represents $v>0$ and blue colour represents $v<0$), and (\emph{d}-\emph{f}) the cross-correlation functions between bottom wall temperature $T_{\text{bottom}}(t)$ and POD mode amplitudes $a_2(t)$ as a function of dimensionless lag time $\tau/T_{\text{period}}$, 
at modulation frequency of (\emph{a},\emph{d}) $f^* = 1$, (\emph{b},\emph{e}) $f^* = 0.1$, (\emph{c},\emph{f}) $f^* = 0.01$, for $Ra = 10^7$ and $Re_b \approx 5623$.}
	\label{fig:POD_Ra1e7}
\end{figure}

To extract the large-scale coherent flow structures, we perform proper orthogonal decomposition (POD) on the turbulent dataset \citep{berkooz1993proper}. 
The POD has been widely employed to study the dynamics of large-scale circulation in convection cells \citep{podvin2015large,castillo2019cessation,soucasse2019proper,xu2021tristable}. 
Specifically, the spatiotemporal flow velocity field $\mathbf{u}(\textbf{x}, t)$ is decomposed into a superposition of empirical orthogonal eigenfunctions $\mathbf{\phi}_i(\mathbf{x})$ and their scalar amplitudes $a_i(t)$ as 
\begin{equation}
	\mathbf{u}(\mathbf{x},t)=\sum_{i=1}^\infty a_i(t) \mathbf{\phi}_i(\mathbf{x}).
\end{equation}
Here, $\mathbf{u}(\mathbf{x},t)=[u(\mathbf{x},t), v(\mathbf{x},t), w(\mathbf{x},t)]^T$ represents the vector field with components $u$, $v$ and $w$,
$\mathbf{\phi}_{i}(\mathbf{x})=[\phi_{i}^{u}(\mathbf{x}), \phi_{i}^{v}(\mathbf{x}), \phi_{i}^{w}(\mathbf{x})]^T$ represents the spatial eigenfunctions (i.e. the POD modes) and $a_{i}(t)$ are the temporal coefficients representing the time-dependent amplitudes of the corresponding modes.
We used at least 900 snapshots to adequately capture the flow structure, ensuring the dominant modes are representative.
At parameters of $Ra = 10^7$ and $Re_b \approx 5623$ (i.e. the corresponding $Ri_{b}=0.445$), we have that the most energetic POD mode corresponds to the streamwise unidirectional shear flow with parallel streamlines pointing along the $x$ direction. 
Then, in figure \ref{fig:POD_Ra1e7}(\emph{a}-\emph{c}), we present the second most energetic POD modes, which are essentially the dominant mode for the fluctuation velocity field. 
Regardless of the temporal modulation frequency of the bottom wall temperature, streamwise-oriented rolls that fill the whole channel height are observed. 
We also examined the time series of mode amplitudes $a_i(t)$  and studied the relationship between wall temperature $T_{\text{bottom}}\left(t\right)=T_{\text{hot}}+2(T_{\text{hot}}-T_{\text{cold}})\sin(2\pi ft)$  and the second POD mode amplitude $a_2(t)$ by calculating their cross-correlation functions as
\begin{equation}
R_{T_{\text{bottom}},a_2}\left(\tau\right)=\frac{\left\langle\left[T_{\text{bottom}}\left(t+\tau\right)-\left\langle T_{\text{bottom}}\right\rangle\right]\left[ a_2(t)-\left\langle a_2 \right\rangle\right]\right\rangle}{\sigma_{T_{\text{bottom}}}\sigma_{a_2}}
\end{equation}
where $\sigma_{T_{\text{bottom}}}$ and $\sigma_{a_2}$ are the standard deviation of $T_{\text{bottom}}$ and ${a_2}$, respectively. 
As shown in figure \ref{fig:POD_Ra1e7}(\emph{d}-\emph{f}), at higher modulation frequency of $f^* = 1$ and $f^* = 0.1$, $T_{\text{bottom}}$ are uncorrelated with ${a_2}$,  indicating that the large-scale rolls are not influenced by changes in wall temperature.
However, at a lower modulation frequency of $f^* = 0.01$, we observe a strong correlation between $T_{\text{bottom}}$ and $a_2$, implying that the strength of the large-scale rolls is significantly affected by temperature modulation on the wall. 
These large-scale rolls are a recognized structural characteristic of atmospheric boundary layers, present in scenarios with mean streamwise flow and exhibiting overall streamwise helical patterns. 
These patterns are responsible for transporting warmer air towards the capping inversion and cooler air towards the ground \citep{jayaraman2021transition}.

\begin{figure}
 	\centerline{\includegraphics[width=0.8\textwidth]{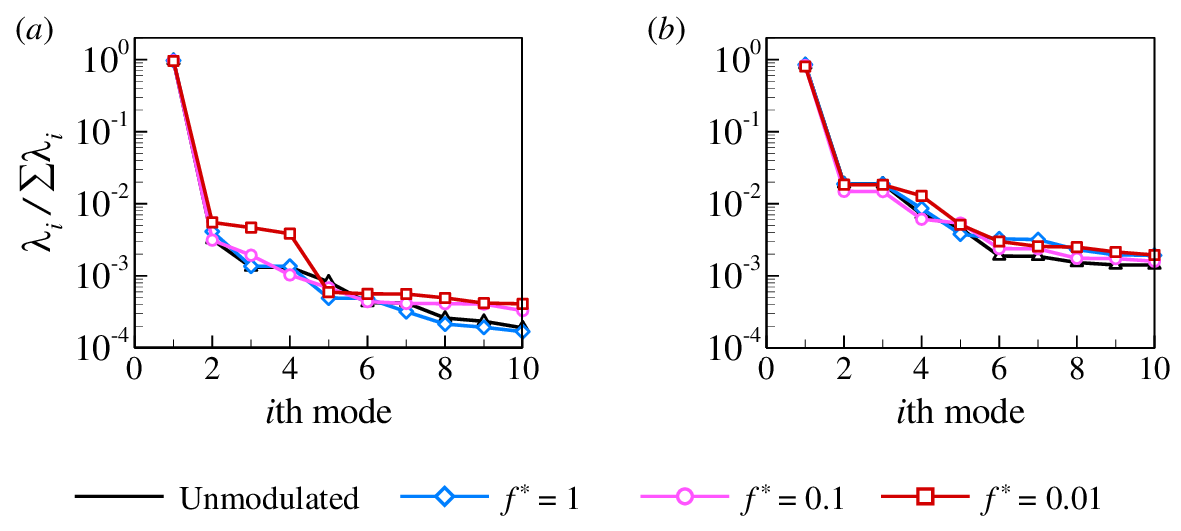}}
 	\caption{The energy contained in each mode, for (\emph{a}) $Ra = 10^{7}$ and (\emph{b}) $Ra = 10^{8}$.}
 	\label{fig:energySpectra}
\end{figure}

We then computed the energy content of the $i$th POD mode $\lambda_i$. 
In figure \ref{fig:energySpectra}, we show the spectra of $\lambda_i$, where each value of $\lambda_i$ is normalized by the total energy $\sum \lambda_{i}$. 
At $Ra = 10^{7}$, the energy content of the first POD mode $\lambda_1$ dominates, accounting for over 95\% of the total energy and representing the mean flow (streamwise unidirectional shear flow). 
The energy content of the second and third POD modes $\lambda_2$ and $\lambda_3$ each contributes approximately 0.1\%-0.6\% of the total energy. 
At a higher Rayleigh number of $Ra=10^8$, with the same bulk Reynolds number of $Re_b \approx 5623$, the energy contained in the first mode reduces to approximately 80\% due to the enhanced effect of buoyancy, while the second and third modes each contribute around 2\% of the total energy. 
These results indicate that both the second and third modes play a role in capturing the flow dynamics.

\begin{figure}
 	\centerline{\includegraphics[width=0.8\textwidth]{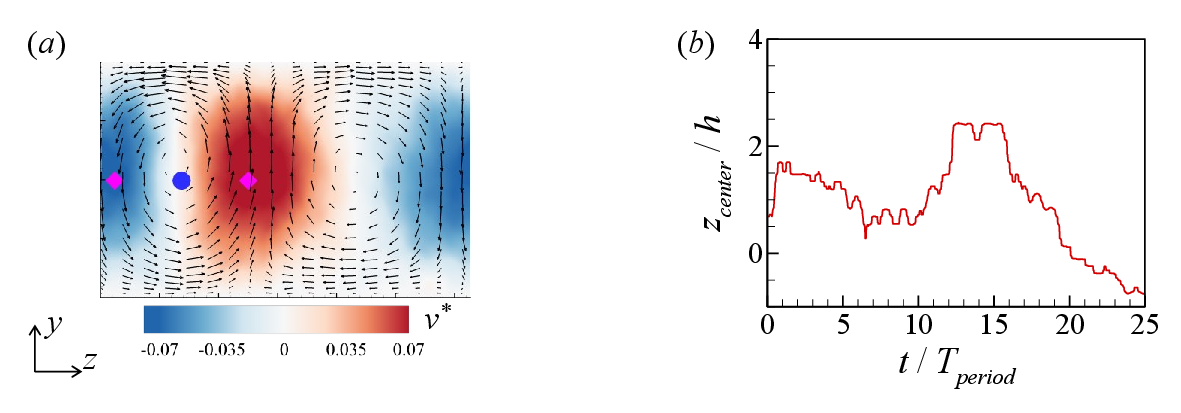}}
 	\caption{(\emph{a}) Typical instantaneous flow field recovered using the second and third POD modes at $Ra=10^{7}$, $Re_{b} \approx 5623$ and $f^*=0.1$. 
 The contour represents the vertical velocity component, and the arrow represents the velocity field of recovered velocity components. 
 The magenta diamonds mark the edges of the rolls and the blue circles mark the roll centre. 
 (\emph{b}) The time series of the position of the roll centre.}
 	\label{fig:POD_recover}
\end{figure}

Due to the periodicity in the wall boundary conditions, the streamwise-oriented rolls are non-stationary and continually move along the spanwise direction. 
To illustrate this, we quantitatively describe the movement of the rolls by tracking their centre. 
Because the axes of these rolls align with the streamwise direction and do not exhibit the complex motion seen in RB convection \citep{vogt2018jump,li2022counter,teimurazov2023oscillatory}, we can track their edges by identifying locations where the vertical velocity component is minimal or maximal. 
The centre of each roll is then determined by calculating the arithmetic mean of these points. 
The POD analysis reveals that not only the second most energetic POD modes but also higher POD modes can represent these rolls. 
For example, at $Ra = 10^7$, $Re_b \approx 5623$ and $f^* = 0.1$, both the second and third POD modes correspond to streamwise-oriented rolls, albeit offset along the spanwise direction. 
Thus, we use both the second and third POD modes to recover the streamwise-oriented roll at $f^* = 0.1$. 
Figure \ref{fig:POD_recover}(\emph{a}) shows the instantaneous recovered flow field at $x = L/2$, representing a pair of counter-rotating streamwise-oriented rolls. 
Here, we mark the edges and centre of one roll to demonstrate the effectiveness of our approach in tracking the roll motion.
When the roll centre exits one side of the domain, we account for the domain's periodicity by wrapping its position to the opposite side, resulting in a continuous trajectory.
In figure \ref{fig:POD_recover}(\emph{b}), we plot the time series of the centre of this roll along the spanwise direction. 
The results suggest that the streamwise-oriented roll moves along the spanwise direction, occasionally appearing to exit from one side of the domain and re-enter from the other.
 
\begin{figure}
 	\centerline{\includegraphics[width=0.8\textwidth]{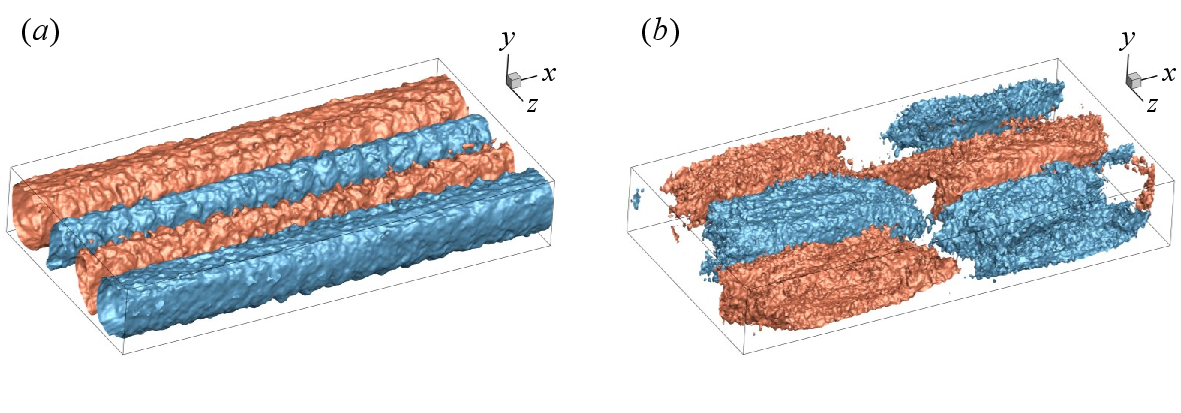}}
 	\caption{The second POD mode in a domain size of $4\pi h \times 2 h \times 2 \pi h$, visualizations of isosurfaces of vertical velocity (red colour represents $v>0$ and blue colour represents $v<0$) at (\emph{a}) $Ra = 10^{7}$, (\emph{b}) $Ra = 10^{8}$, with $f^* = 0.1$.}
 	\label{fig:POD_largeDomain}
\end{figure}

A smaller domain may restrict the formation of large structures due to the imposed periodic boundary conditions and limited spatial extent \citep{stevens2024wide};  thus we conducted additional simulations in a larger domain ($4\pi h \times 2h \times 2 \pi h$). 
In this expanded domain, at $Ra=10^7$ and $Re_b \approx 5623$ (i.e. the corresponding $Ri_b=0.445$), we observed two pairs of counter-rotating straight rolls (see figure \ref{fig:POD_largeDomain}\emph{a}), which align with expectations based on domain scaling.
In other words, doubling the horizontal extent resulted in two roll pairs, compared with one pair in the smaller domain. 
We then examined the second POD modes in this large domain at $Ra=10^8$ and $Re_b \approx 5623$ (i.e. the corresponding $Ri_b=4.45$).
As shown in figure \ref{fig:POD_largeDomain}(\emph{b}), the isosurfaces reveal variations in the vertical velocity, with alternating regions of upward (positive) and downward (negative) flow. 
The higher Rayleigh number induces stronger buoyancy forces and increased thermal driving, which leads to the breakdown of coherent rolls into smaller, more chaotic structures. 
While these structures remain somewhat aligned in the streamwise direction, they appear increasingly fragmented, indicating a transition towards a more convection-dominated state. 
Accordingly, we refer to these structures at $Ra = 10^8$ as fragmented streamwise-oriented rolls.

\subsection {Long-time-averaged statistics}

\begin{figure}
	\centerline{\includegraphics[width=0.8\textwidth]{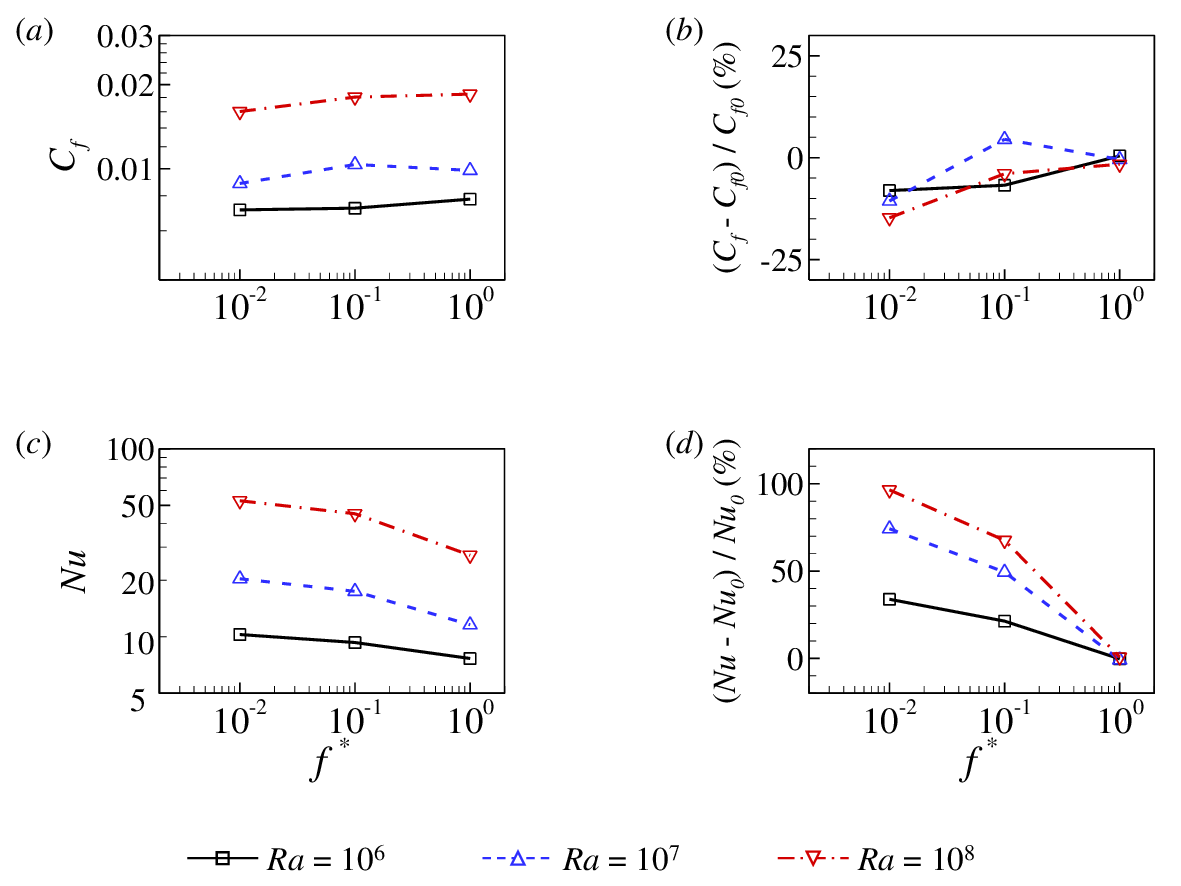}}
	\caption{(\emph{a}) Friction coefficient, (\emph{b}) values of $C_f/C_{f0}-1$, (\emph{c}) Nusselt number and (\emph{d}) values of $Nu/Nu_0-1$, as functions of $f^*$ for various $Ra$. 
Here, $C_{f0}$ and $Nu_0$ are the friction coefficient and Nusselt numbers without wall temperature modulation, respectively.}
	\label{fig:Cf_Nu}
\end{figure}

To study changes induced on the mean flow and temperature by the wall temperature modulation, we now focus on the long-time-averaged statistics. 
In figure \ref{fig:Cf_Nu}, we present statistics of aerodynamic drag and heat transfer as a function of modulation frequency for various Rayleigh numbers, which is a topic of interest in meteorology and engineering \citep{yerragolam2022small,yerragolam2024scaling}. 
In figure  \ref{fig:Cf_Nu}(\emph{a}), we examine the aerodynamic drag in terms of friction coefficients ($C_f$) at the bottom wall, which is calculated as $C_f=2\langle \tau_w \rangle/(\rho u_b^2)$. 
The increased drag due to the emission of plumes in thermal field is consistent with that reported by \citet{scagliarini2015law}, \citet{pirozzoli2017mixed} and \citet{howland2024turbulent}. 
With wall temperature modulation, we observe that aerodynamic drag is not very sensitive to the wall temperature modulation frequency. 
Figure \ref{fig:Cf_Nu}(\emph{b}) further visualizes the relative changes of $C_f$ with wall temperature modulation, showing that the variation in $C_f$ is less than 15\%. 
In figure  \ref{fig:Cf_Nu}(\emph{c}), we examine the heat transfer efficiency in terms of the Nusselt number ($Nu$), which is calculated as $Nu=\sqrt{RaPr/Ri_{b}}\left\langle v^* T^*\right\rangle_{V,t}+1 $. 
The $Nu$ at the highest modulation frequency is the same as that without wall temperature modulation. 
With the decrease in modulation frequency, we observe enhanced heat transfer efficiency for all Rayleigh numbers. 
Previously, in pure RB convection, \citet{yang2020periodically} reported a regime where the modulation is too fast to affect $Nu$, a regime where $Nu$ increases with decreasing $f^*$ and a regime where $Nu$ decreases with further decreasing $f^*$. 
Our results in mixed PRB convection cover the first two regimes reported in pure RB convection, yet we did not further explore slower frequency due to the high computational cost in the 3-D simulation. 
Figure \ref{fig:Cf_Nu}(\emph{d}) further visualizes that among the three Rayleigh numbers, $Ra = 10^8$ results in the largest magnitude of heat transfer efficiency, up to 96\%. 
At high Rayleigh numbers, convective patterns are more vigorous, and wall temperature modulation has a more disruptive effect. 
Higher frequencies likely cause more frequent disturbances in the boundary layers, reducing heat transfer efficiency by breaking up coherent thermal plumes.

 \begin{figure}
	\centerline{\includegraphics[width=0.95\textwidth]{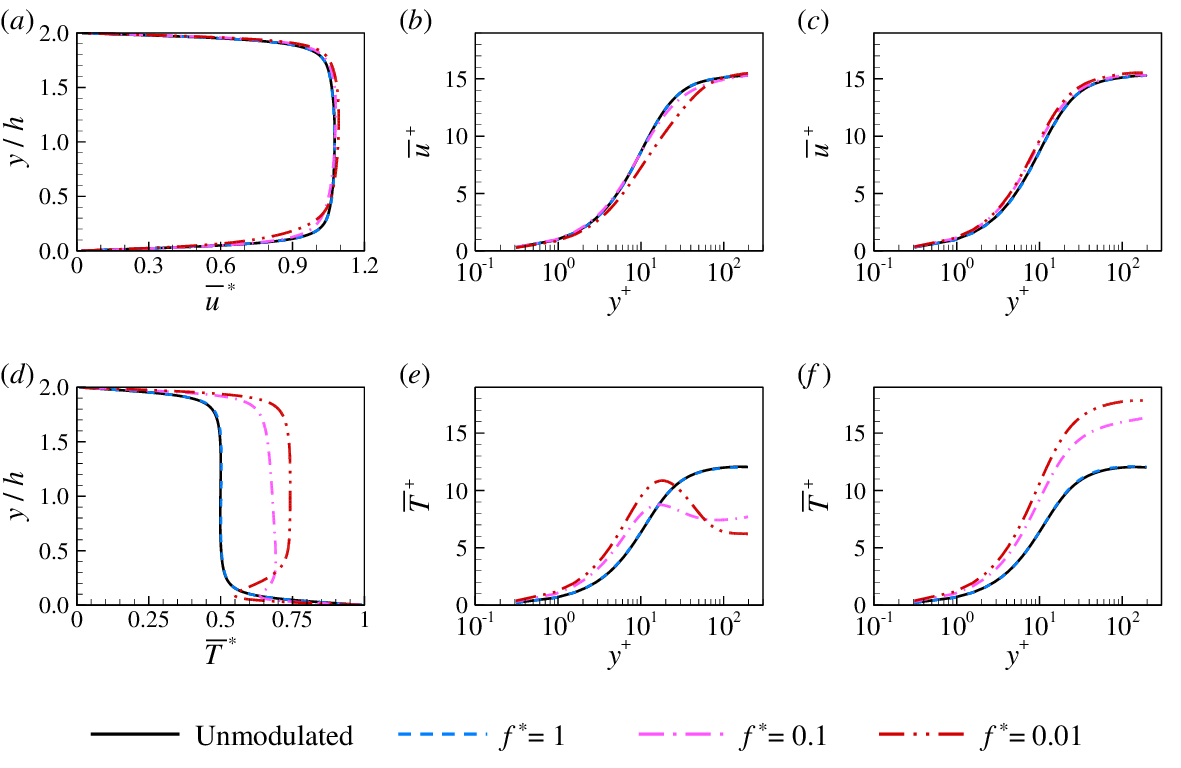}}
	\caption{Mean profiles of (\emph{a}-\emph{c}) streamwise velocity and (\emph{d}-\emph{f}) temperature 
(\emph{a},\emph{d}) along the whole channel height, (\emph{b},\emph{e}) along the bottom half-height and (\emph{c},\emph{f}) along the top half-height, for $Ra = 10^7$ and $Re_b \approx 5623$.}
	\label{fig:meanProfile}
\end{figure}

In figure \ref{fig:meanProfile}, we show the mean profiles for streamwise velocity and temperature across the channel height to illustrate the effects of wall temperature modulation on the mean flow. 
We denote the mean velocity and mean temperature by an overbar, and the fluctuating quantities by a prime; thus we have $u=\overline{u}+u'$ and $T=\overline{T}+T'$.
Consistent with our observations on flow organization, these mean profiles are similar between scenarios without wall temperature modulation and those with the highest modulation frequency ($f^* = 1$). 
For the mean streamwise velocity profile, it is symmetric about the mid-plane $y = h$ at the highest frequency of $f^* = 1$; 
however, slight asymmetry is observed at lower frequencies of $f^* = 0.1$ and 0.01 (see figure \ref{fig:meanProfile}\emph{a}). 
This asymmetry is also evident from table \ref{tab:Numerical details}, which shows the relative differences in friction Reynolds numbers. 
To highlight the modulation effect in the near-wall regions, we replot the velocity data using wall scaling in figures \ref{fig:meanProfile}(\emph{b}) and \ref{fig:meanProfile}(\emph{c}). 
Here, the superscript '+' indicates normalization using the kinematic viscosity $\nu$ and the friction velocity $u_{\tau}$ in wall units. 
The distances from the walls are then given by $y^+=yu_{\tau}/\nu$.
We can see that at lower frequencies, the average streamwise velocity $\overline{u}^{+}=\overline{u}/u_{\tau}$ decreases near the bottom wall (see figure \ref{fig:meanProfile}\emph{b}) and increases near the top wall (see figure \ref{fig:meanProfile}\emph{c}). 
As for the mean temperature profile, at $f^* = 1$, the temperature decreases monotonically from the bottom wall, stabilizing at a constant value of the arithmetic mean temperature $T^* = 0.5$ in the bulk region, and is antisymmetric about the mid-plane (see figure \ref{fig:meanProfile}\emph{d}). 
This pattern recovers the canonical RB convection profile. 
However, at $f^* = 0.1$ and 0.01, deviations from the profiles without wall modulation are evident. 
The temperature first decreases and then increases near the bottom wall, until it reaches a plateau in the bulk region. 
Similar to findings in RB convection \citep{yang2020periodically}, our results suggest that at lower modulation frequencies, the bulk temperature increases compared with cases without wall temperature modulation.
We also replot the temperature data using wall scaling, as shown in figures \ref{fig:meanProfile}(\emph{e}) and \ref{fig:meanProfile}($f$). 
For a straightforward comparison between the heating and cooling sides, we calculate the average temperature as $\overline{T}^{+}=\left|\overline{T}-\langle T_{\text{wall}}\rangle\right|/T_{\tau}$.
Here, $\langle T_{\text{wall}}\rangle$ is the mean temperature at either the bottom wall or top wall,
and the friction temperature $T_{\tau} = Q/u_{\tau}$  is used to normalize the average temperature, where $Q$ is total vertical heat flux, calculated as $Q = (\alpha \Delta_{T}/H) Nu$.
At $f^* = 0.1$ and 0.01, we observe a peak in the $\overline{T}^{+}$ profile at $y^+ \approx17 $ (see figure \ref{fig:meanProfile}\emph{e}) due to the formation of thermal Stokes layers by the oscillatory wall temperature \citep{yang2020periodically}; 
however, such a peak is absent at a higher frequency of $f^* = 1$, as well as near the top wall where the wall temperature is constant (see figure \ref{fig:meanProfile}$f$).

\begin{figure}
	\centerline{\includegraphics[width=0.8\textwidth]{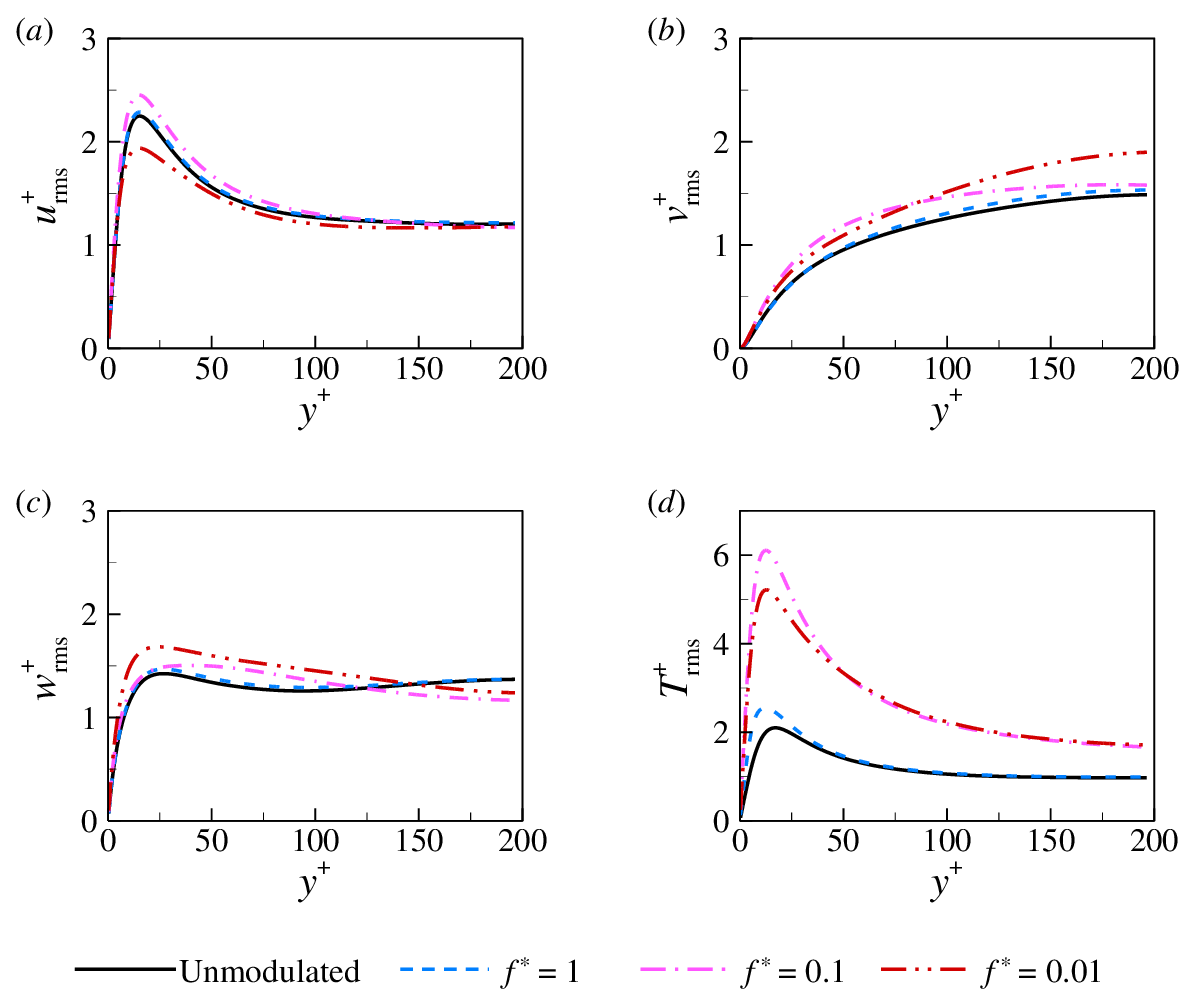}}
	\caption{The r.m.s. fluctuation of (\emph{a}) streamwise velocity, (\emph{b}) wall-normal velocity, (\emph{c}) spanwise velocity and (\emph{d}) temperature along the bottom half-height of the channel, for $Ra = 10^7$ and $Re_b \approx 5623$.}
	\label{fig:rmsProfile}
\end{figure}

To examine the influence of wall-temperature modulation on turbulence quantities, in figure \ref{fig:rmsProfile}, we present the root-mean-square (r.m.s.) fluctuation of velocity components and temperature along the lower half-height of the channel. 
The peak of the r.m.s. streamwise velocity profile consistently occurs at $y^+\approx15 $ (see figure \ref{fig:rmsProfile}\emph{a}), where turbulent eddies are most active, similar to observations in turbulent channel flow without convection \citep{moser1999direct}. 
The wall-normal velocity ﬂuctuation increases throughout the half-channel height due to wall temperature modulation, becoming comparable to or even larger than the streamwise component (see figure \ref{fig:rmsProfile}\emph{b}). 
These fluctuations, sensitive to buoyancy effects, suggest that lower-frequency thermal modulations enhance buoyancy-driven turbulence. 
The spanwise velocity fluctuation also increases in the near-wall region with wall temperature modulation (see figure \ref{fig:rmsProfile}\emph{c}). 
This increase in wall-normal and spanwise velocity fluctuation implies that fluid columns rising from the bottom wall activate cross-stream eddies near the wall. 
As for the r.m.s. of the temperature (see figure \ref{fig:rmsProfile}\emph{d}), lower-frequency wall temperature modulations ($f^* = 0.01$) yield higher peaks in temperature fluctuations near the wall, indicating a more unstable thermal boundary layer with larger eddies under slower temperature modulation. 
In contrast, higher frequencies ($f^* = 1$) appear to stabilize the thermal boundary layer, leading to less pronounced temperature fluctuations. 
 
\begin{figure}
	\centerline{\includegraphics[width=0.8\textwidth]{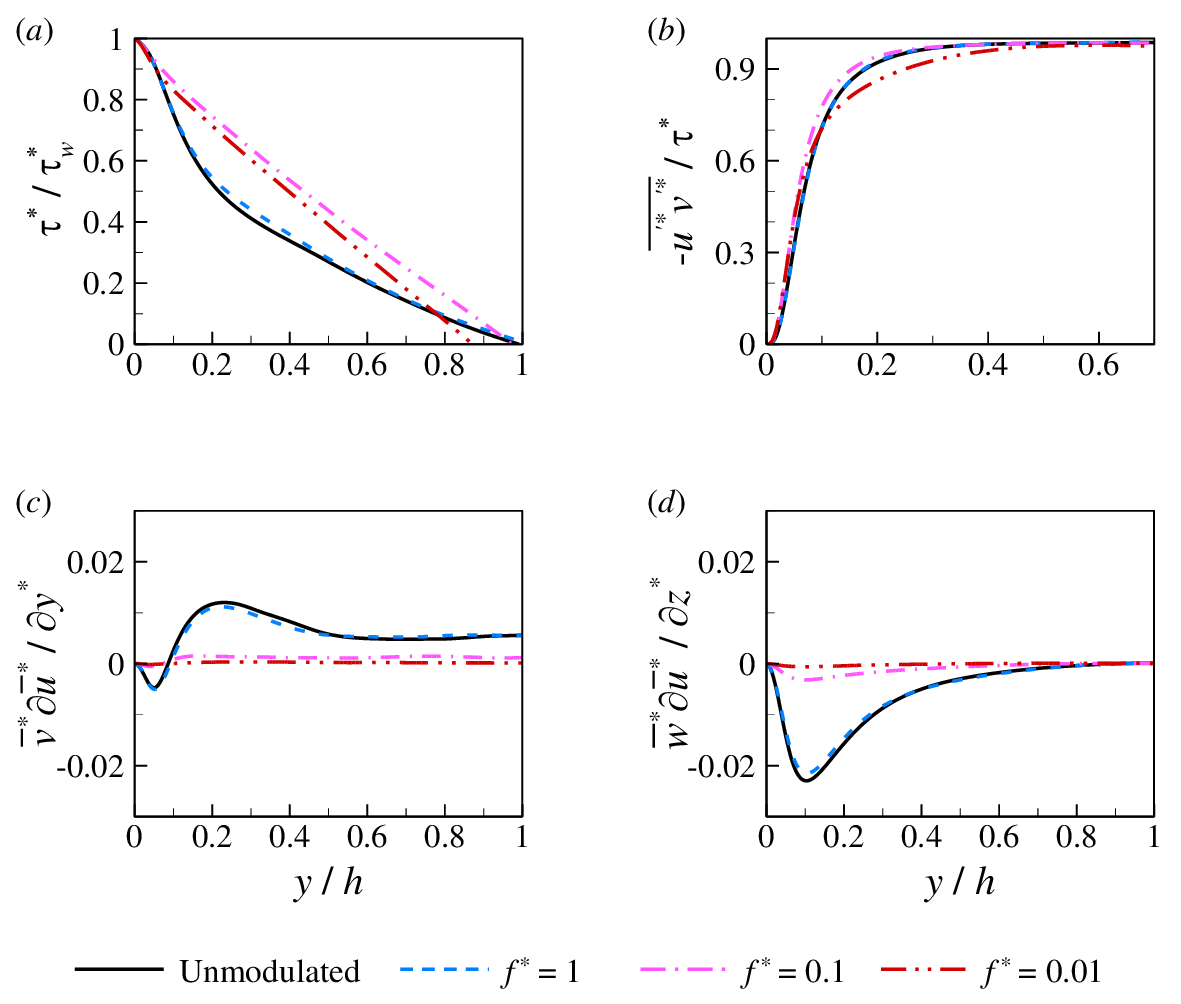}}
	\caption{Profile of (\emph{a}) total shear stress, (\emph{b}) percentage of Reynolds stress to the total shear stress, (\emph{c}) convective term component $\overline{v}^*\partial\overline{u}^*/\partial y^{*} $ and (\emph{d}) convective term component $ \overline{w}^*\partial\overline{u}^*/\partial z^{*}$ along the channel half-height, for $Ra = 10^7$ and $Re_b \approx 5623$.}
	\label{fig:stressProfile}
\end{figure}

Wall temperature modulation also influences the wall-normal behaviour of momentum flux. 
Because the mean flow is predominantly in the axial direction for our investigated flow parameters, we plot the distribution of shear stress $\tau(y)=\rho \nu d\overline{u}/dy-\overline{u^{\prime}v^{\prime}}$ in terms of dimensionless function $\tau^* / \tau_w^*$ across the lower-half of the channel height (from $y=0$ to $y=h$) in figure \ref{fig:stressProfile}(\emph{a}), where $\tau_w^*$ is the dimensionless wall shear stress. 
In figure \ref{fig:stressProfile}(\emph{b}), we further show the contribution of Reynolds stress $-\overline{u^{\prime*}v^{\prime*}} $ to the total shear stress $\tau^*$.
At a higher frequency of $f^* = 1$, the total shear stress exhibits a nonlinear distribution, whereas at lower frequencies of $f^* = 0.1$ and $f^* = 0.01$, it decreases linearly along the channel’s half-height. 
The linear distribution of $\tau^*/\tau_w^*$  at lower frequencies resembles the shear stress profile observed in a pure turbulent channel without convection. 
To further explain the deviation from linear total shear stress at higher frequencies, we examine the mean-momentum equation along the streamwise direction within the mixed convection channel:
\begin{equation}\label{x_momentumequation}
    \begin{split}
 \overline{u}^*\frac{\partial\bar{u}^*}{\partial x^*}
+\overline{v}^*\frac{\partial\bar{u}^*}{\partial y^*}
+\overline{w}^*\frac{\partial\bar{u}^*}{\partial z^*} 
& =-\frac{\partial\overline{P}^*}{\partial x^*}
+\frac{1}{Re_{b}} \left(\frac{\partial^2\overline{u}^*}{\partial x^{*2}}
         +\frac{\partial^2\overline{u}^*}{\partial y^{*2}}
         +\frac{\partial^2\overline{u}^*}{\partial z^{*2}}\right) \\
& -\left(\frac{\partial\overline{u^{\prime*}u^{\prime*}}}{\partial x^*}
      +\frac{\partial\overline{u^{\prime*}v^{\prime*}}}{\partial y^*}
      +\frac{\partial\overline{u^{\prime*}w^{\prime*}}}{\partial z^*}\right)
+\overline{f_b}^* \\
    \end{split}
\end{equation}
Here, the terms for $\partial\left(\cdot\right)/\partial x^*$  and $ \partial^2(\cdot)/\partial x^{*2}$ are near zero in the fully developed region, where velocity statistics no longer vary with the streamwise direction, as confirmed by our DNS results. 
Due to the detachment of thermal plumes from the wall and their horizontal spreading, the homogeneous condition along the spanwise direction may be violated, making the terms $ \partial(\cdot)/\partial z^*$  and $ \partial^2(\cdot)/\partial z^{*2}$ non-zero near the wall. 
However, we numerically verified (not shown here for simplicity) that the magnitudes of $(1/Re_b)\partial^2\overline{u}^*/\partial z^{*2}$ and $\partial\overline{u^{\prime*}w^{\prime*}}/\partial z^*$ are three orders smaller than the other terms, so they can be neglected. 
Thus, (\ref{x_momentumequation}) can be rewritten as 
\begin{equation}
\frac{\partial}{\partial y^*}\left(\frac{1}{Re_b}\frac{\partial \overline{u}^*}{\partial y^*}-\overline{u^{\prime*}v^{\prime*}}\right)
=\overline{v}^*\frac{\partial\overline{u}^*}{\partial y^*}+\overline{w}^*\frac{\partial\overline{u}^*}{\partial z^*}-\overline{f_b}^*
\ \ \ \Rightarrow \ \ \
\frac{d\tau^*}{dy^*}=\overline{v}^*\frac{\partial\overline{u}^*}{\partial y^*}+\overline{w}^*\frac{\partial\overline{u}^*}{\partial z^*}-\overline{f_b}^*
\end{equation}
At a higher frequency of $f^* = 1$, the convection effects are more pronounced for the mean flow, resulting in finite values for the components $\overline{v}^*\partial\overline{u}^*/\partial y^*$ and $ \overline{w}^*\partial\overline{u}^*/\partial z^*$, as shown in figures \ref{fig:stressProfile}(\emph{c}) and \ref{fig:stressProfile}(\emph{d}). 
At lower frequencies of $f^* = 0.1$ and $0.01$, the convection effects are substantially reduced for the mean flow, making $ \overline{v}^*\partial\overline{u}^*/\partial y^*$ and $\overline{w}^*\partial\overline{u}^*/\partial z^*$ close to zero, restoring conditions similar to those for a pure turbulent channel. 
This reduced convection for the mean flow is mainly attributed to the formation of a stably stratified layer near the bottom wall, which acts as a thermal blanket, effectively suppressing buoyancy forces.

\begin{figure}
	\centerline{\includegraphics[width=0.8\textwidth]{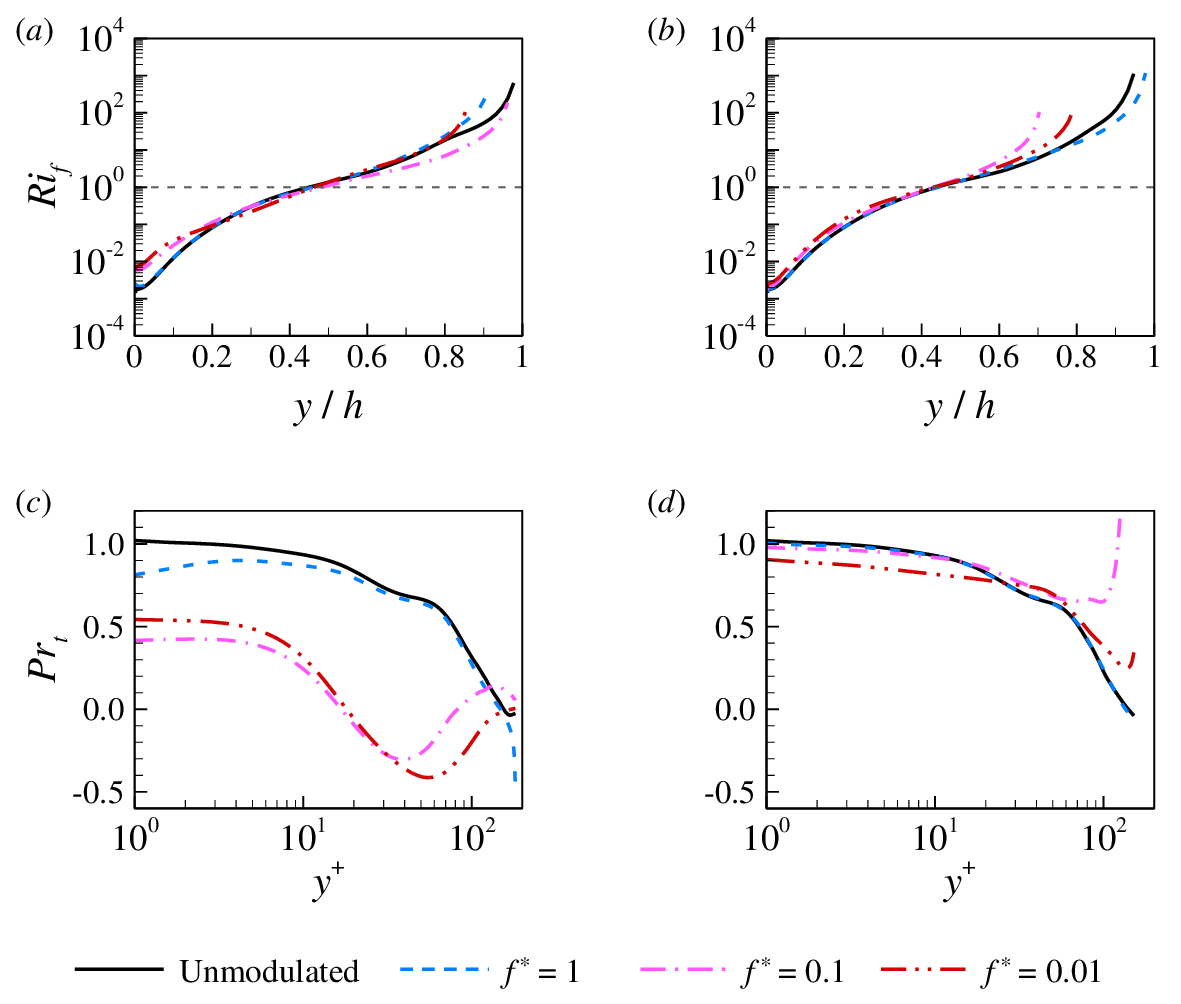}}
	\caption{Profiles of (\emph{a},\emph{b}) flux Richardson number and (\emph{c},\emph{d}) turbulent Prandlt number, along the (\emph{a},\emph{c}) the bottom half-channel and (\emph{b},\emph{d}) the top half-channel, respectively, for $Ra = 10^7$ and $Re_b \approx 5623$.}
	\label{fig:Rif_Prt}
\end{figure}

We quantify the local relative dynamic importance of buoyancy compared with friction using the flux Richardson number, which is calculated as \citep{pirozzoli2017mixed}
\begin{equation}
Ri_f=\frac{-\beta g\overline{\nu^{\prime}T^{\prime}}}{\overline{u^{\prime}\nu^{\prime}}d\bar{u}/dy}
\end{equation}
In figures \ref{fig:Rif_Prt}(\emph{a}) and \ref{fig:Rif_Prt}(\emph{b}), we plot $Ri_f$ along the bottom and top halves of the channel height, respectively. 
Regardless of the wall modulation frequency, the near-wall region ($y \leq 0.2h$) is dominated by shear. 
However, as we move farther away from the wall, convection begins to emerge and eventually dominates in the bulk region. 
We also quantify the ratio of turbulent momentum to thermal diffusivity via the turbulent Prandlt number $Pr_t$, which is frequently used in modelling turbulent heat transfer. 
For simple shear flows, the Reynolds analogy suggests that $Pr_t$ is of the order of unity \citep{kays1994turbulent}.  
Using DNS results, we can calculate $Pr_t$ as
\begin{equation}
Pr_t=\frac{\nu_t}{\alpha_t}=\frac{\overline{u'v'}}{\overline{v'T'}}\frac{d\overline{T}/dy}{d\overline{u}/dy}
\end{equation}
Here,  $\nu_t$ is turbulent viscosity and $\alpha_t$ is the thermal eddy diffusivity. 
On the bottom side (see figure \ref{fig:Rif_Prt}\emph{c}), at a high frequency of $f^* = 1$ and the unmodulated case, $Pr_t$ is close to unity in the near-wall region; 
however, at lower modulation frequencies, $Pr_t$ deviates from unity significantly because the temperature is more efficiently transported compared with momentum. 
On the top side (see figure \ref{fig:Rif_Prt}\emph{d}), $Pr_t$ approximates unity near the wall, yet it exhibits a large variation in the bulk region. 
These features of the $Pr_t$ distribution present challenges in turbulence modelling, where a constant value is often assumed.

\subsection {Phase-averaged statistics }

\begin{figure}
	\centerline{\includegraphics[width=0.8\textwidth]{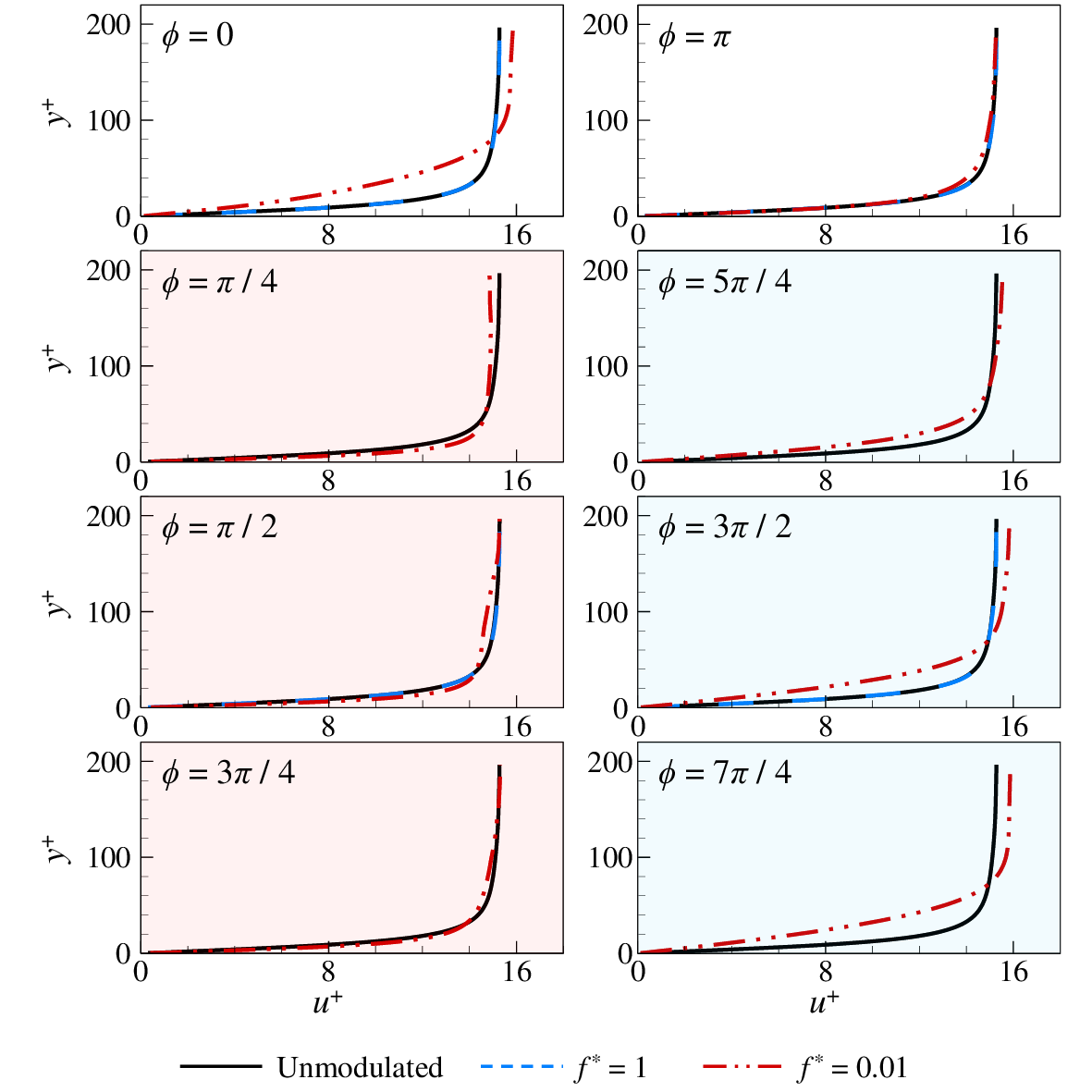}}
	\caption{Phase-averaged streamwise velocity profiles $\left\langle u^{+}(y^{+})\right\rangle_\phi $  along the bottom half-height for $Ra = 10^7$ and $Re_b \approx 5623$. 
For the case without wall temperature modulation, we present the long-time-averaged profiles to guide the eye. 
}
	\label{fig:phaseAverage_U}
\end{figure}

Because the wall temperature modulation inherently induces unsteady flow and temperature evolution, we now focus on the phase-averaged statistics. 
We present the variability of the physical quantities during different phases of the flow cycle and how they are affected by the wall temperature modulation. 
In the following, we denote with angle brackets $\left\langle\cdot\right\rangle $ all quantities that have been phase-averaged. 
We first calculate the phase-averaged streamwise velocity $\left\langle u^{+}(\mathbf{x})\right\rangle_\phi$ and plot its profile $\left\langle u^{+}\left(y^{+}\right)\right\rangle_\phi=\left[\int_0^{L}\int_0^{W}\left\langle u^{+}\left(\mathbf{x}\right)\right\rangle_\phi dzdx\right]/\left(LW\right)$ along the bottom-half of the channel height, as shown in figure \ref{fig:phaseAverage_U}. 
At a high frequency of $f^* = 1$, the flow maintains a velocity profile consistent with that under constant wall temperature. 
However, at a lower frequency of $f^* = 0.01$, the velocity profiles deviate from those without wall temperature modulation, particularly in the near-wall region where the thermal boundary layer causes variations in the velocity gradients. 
An interesting observation is that during the cooling phase (e.g. at a phase angle of $\phi = 5\pi/4, 3\pi/2, 7\pi/4$), the streamwise velocity near the wall significantly decreases. 
This occurs because the turbulence intensity near the bottom wall is greatly weakened by the cooling. 
As a result, the reduced turbulence results in lower surface drag on the fluid above, leading to an acceleration of the flow and a higher plateau away from the wall compared with the constant wall temperature condition. 
This trend aligns with the dynamics of the nocturnal low-level jet in atmospheric flow. 
At night, surface cooling creates stable conditions and temperature inversions, which suppress turbulence near the ground. 
This suppression reduces surface drag and allows for the formation of nocturnal jets characterized by higher wind speeds above the cooled surface layer. 
Understanding the behaviour of nocturnal jets is crucial for sandstorms, air pollution, wind energy utilization, and aviation safety \citep{liu2014advances}.

\begin{figure}
	\centerline{\includegraphics[width=0.8\textwidth]{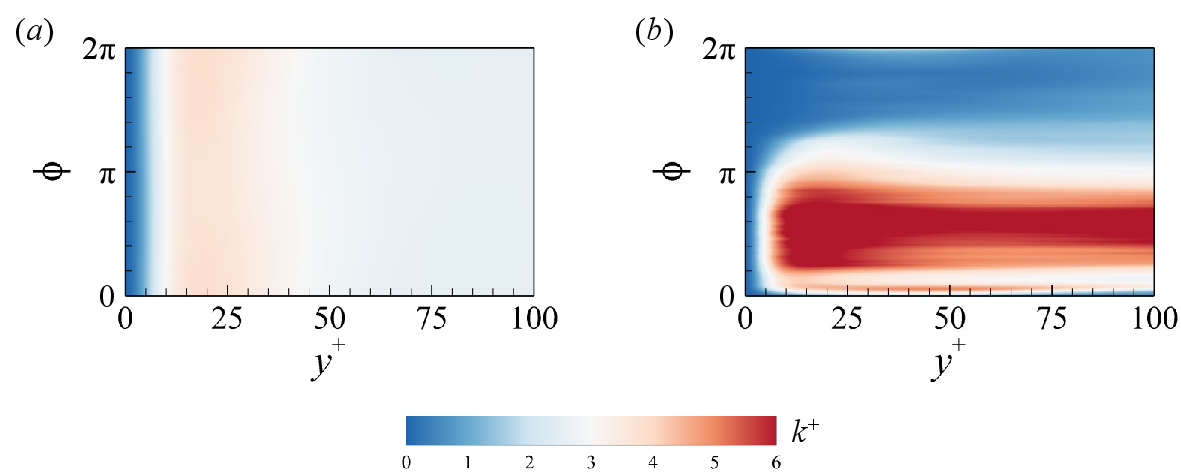}}
	\caption{Space-phase diagrams of turbulent kinetic energy (TKE) along the bottom half-height for $Ra = 10^7$ and $Re_b \approx 5623$: (\emph{a}) $f^* = 1$ and (\emph{b}) $f^* = 0.01$.}
	\label{fig:TKE}
\end{figure}

The behaviour of velocity distribution can be further understood by studying the phase-averaged TKE as $k^+=\langle {u_i^{\prime+}u_i^{\prime+}}\rangle_{\phi}/2$. 
Here, the subscript $i$ is a dummy index, and the superscript ($^\prime$) denotes the fluctuation part of an instantaneous flow variable. 
In figure \ref{fig:TKE}, we show the space-phase diagram of turbulence intensities. 
At a high frequency of $f^* = 1$ (see figure \ref{fig:TKE}\emph{a}), we can infer that the perturbation is relatively intense in the region $10 \leq y^+\leq 40$  across different phases, and such consistent TKE leads to phase-averaged velocity profiles that are insensitive to phase variations. 
At a low frequency of $f^* = 0.01$ (see figure \ref{fig:TKE}\emph{b}), there is phase-dependent variation in TKE diagrams, and turbulence intensity is strengthened and reduced alternately in one cycle. 
During heating phases, the flow behaves as unstable stratification, and TKE exhibits significant peaks in the range of $ y^+ \geq 10 $; 
during cooling phases, the flow behaves similarly to stable stratification, TKE decreases to almost zero, and turbulence intensities are suppressed. 
Such reduced TKE results in lower velocity close to the wall and higher velocity aloft, as previously observed in figure \ref{fig:phaseAverage_U}.
We examine the phase-averaged TKE equation of incompressible mixed convection, which is written as 

\begin{equation}
\begin{aligned}
\frac{\partial k^+}{\partial t^+} 
+ \langle u_j^+ \rangle_\phi \partial_{j^+} k^+ 
&= \underbrace{-\langle u_i^{\prime+} u_j^{\prime+} \rangle_\phi \partial_{j^+} \langle u_i^+ \rangle_\phi}_{\langle P \rangle} \\
&\quad 
\underbrace{- \partial_{j^+}\langle P^{\prime+} u_j^{\prime+} \rangle_\phi}_{\langle \Pi \rangle} 
\underbrace{- \frac{1}{2}\partial_{j^+} \langle u_i^{\prime+} u_i^{\prime+} u_j^{\prime+} \rangle_\phi}_{\langle T \rangle} 
+ \underbrace{\partial_{j^+}\partial_{j^+} k^+}_{\langle D \rangle}  \\
&\quad 
\underbrace{- \langle (\partial_{j^+} u_i^{\prime+})^2 \rangle_\phi}_{\langle \varepsilon \rangle} 
+ \underbrace{\frac{Ra}{Pr(2Re_\tau)^3}\langle T^{\prime*} v^{\prime+} \rangle_\phi}_{\langle B \rangle}
\end{aligned}
\end{equation}

\begin{figure}
	\centerline{\includegraphics[width=0.8\textwidth]{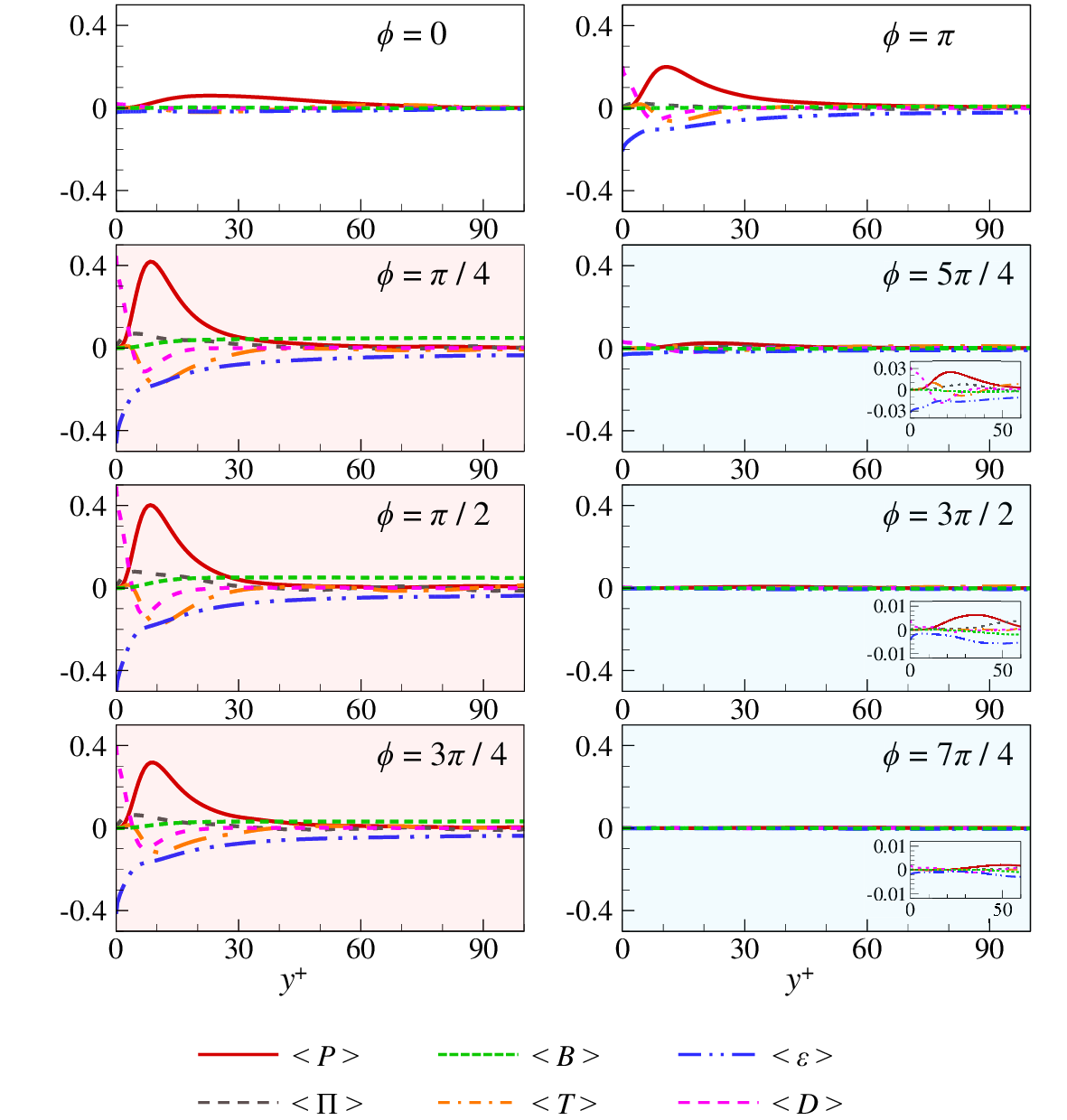}}
	\caption{Phase-averaged TKE budgets along the bottom half-height for $Ra = 10^7$, $Re_b \approx 5623$ and $f^* = 0.01$.}
	\label{fig:TKE_budget}
\end{figure}

In the above, the terms $\langle P \rangle$ and $\langle B \rangle$ represent the production by shear and buoyancy, respectively;
the terms $\langle \Pi \rangle$, $\langle T \rangle$ and $\langle D \rangle$ represent turbulent diffusion by pressure-velocity fluctuations, velocity fluctuations and viscous diffusion, respectively;
and the term $\langle \varepsilon \rangle$ represents dissipation.
In figure \ref{fig:TKE_budget}, we show the TKE budgets for $Ra = 10^7$, $Re_b \approx 5623$ and $f^* = 0.01$. 
The shear-induced and buoyancy-induced TKE productions are strongly influenced by the wall temperature modulation in both amplitude and peak position. 
Specifically, during the heating phase (e.g. at a phase angle of $\phi = \pi/4, \pi/2, 3\pi/4$), the TKE production is dominated by shear in the near-wall region, while it is dominated by buoyancy in the bulk region. 
During the cooling phase (e.g. at a phase angle of $\phi = 5\pi/4, 3\pi/2, 7\pi/4$), the TKE productions are all near zero along the whole channel height. 
For other terms of the budget, including the velocity-pressure gradient, the viscous diffusion and the turbulent transport, the in-cycle variation is also evident during the heating phase in the near-wall region but shows minor variation during the cooling phase (see the insets in figure \ref{fig:TKE_budget}).

\begin{figure}
	\centerline{\includegraphics[width=0.8\textwidth]{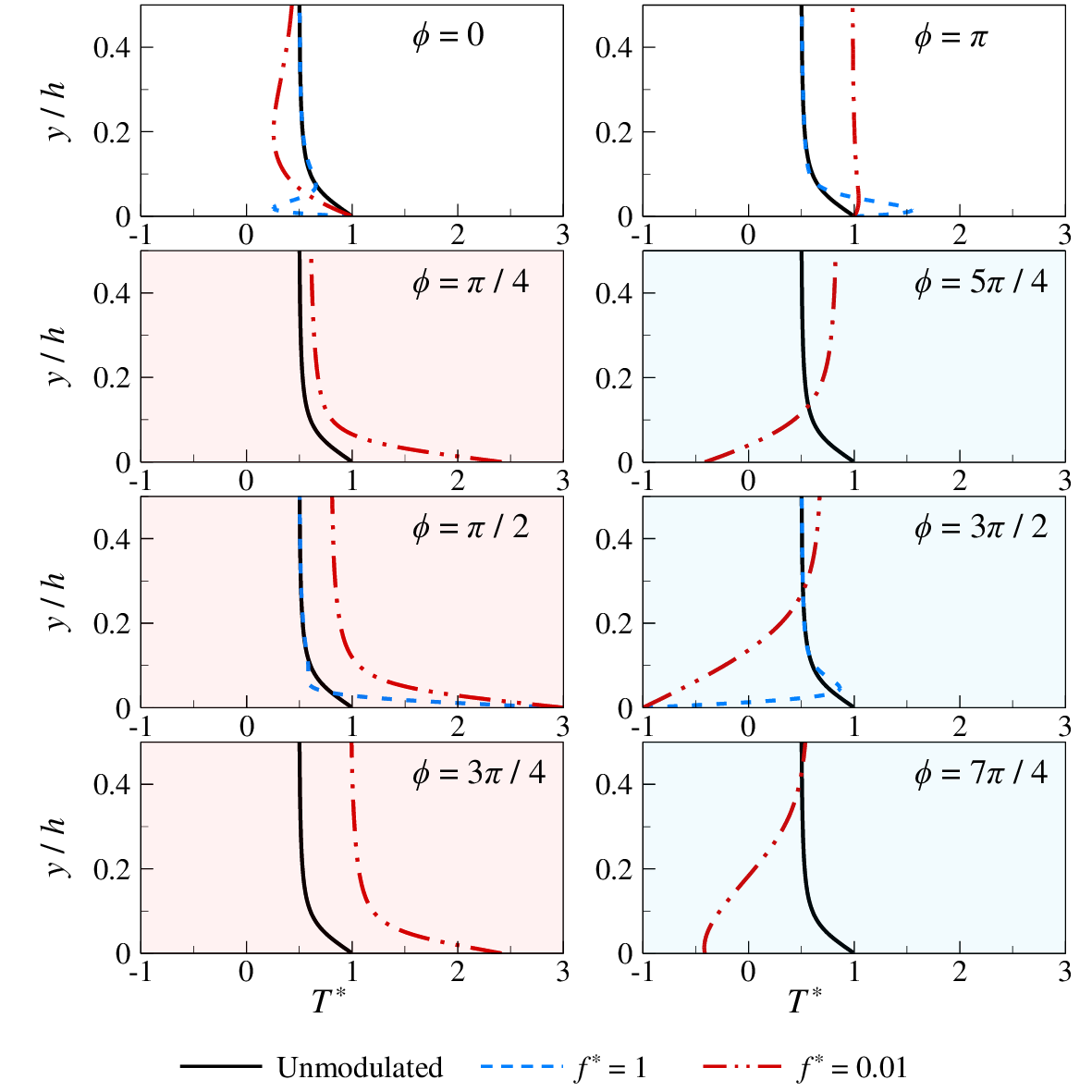}}
	\caption{Phase-averaged temperature profiles $\left\langle T^*(y^*)\right\rangle_\phi $ along the bottom-half channel height for $Ra = 10^7$ and $Re_b \approx 5623$.}
	\label{fig:phaseAverage_T}
\end{figure}

We then analyse the phase-averaged temperature $\left\langle T^* \left(\mathbf{x}\right)\right\rangle_{\phi}$ , and plot its profile $\left\langle T^* \left(y^*\right)\right\rangle_\phi=\left[\int_0^{L}\int_0^{W}\left\langle T^* \left(\mathbf{x} \right)\right\rangle_\phi dzdx\right]/\left(LW\right)$ along the bottom half-channel height, as shown in figure \ref{fig:phaseAverage_T}. 
At a high frequency of $f^* = 1$, the temperature profiles deviate from those without wall temperature modulation only in the region of $y < 0.1h$. 
Farther away from the wall, the influence of the modulation is limited. 
At a lower frequency of $f^* = 0.01$, both the near-wall and bulk temperatures are influenced by the modulation. During the heating phase (e.g. at a phase angle of $\phi = \pi/4, \pi/2, 3\pi/4$), the temperature decreases upward in the boundary layer, resulting in an unstably stratified flow. 
During the cooling phase (e.g. at a phase angle of $\phi = 5\pi/4, 3\pi/2, 7\pi/4$), the temperature increases upward in the boundary layer, forming a statistically stable boundary layer. 
This stabilization suppresses turbulence and reduces mixing, leading to the formation of a residual layer above the stable boundary layer that retains the thermal stratification from the previous cycle. 
In meteorology, this residual layer is analogous to the one that retains the adiabatic lapse rate from the previous day, as described by \citet{stull2015practical}.

\begin{figure}
	\centerline{\includegraphics[width=0.95\textwidth]{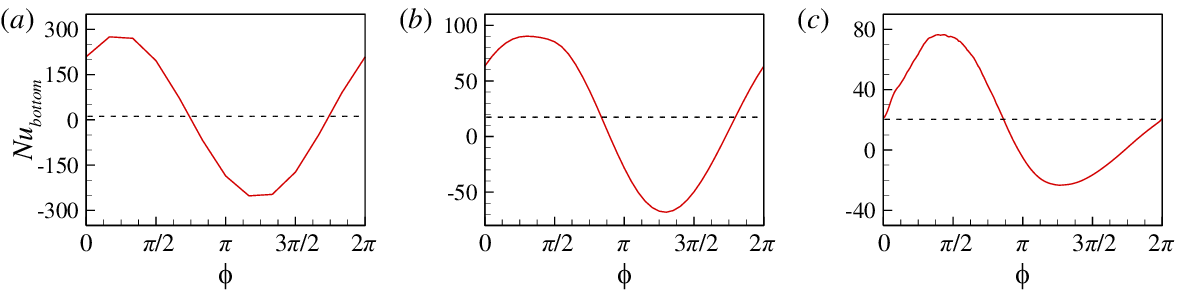}}
	\caption{Phase-averaged dimensionless heat flux in terms of Nusselt number at the bottom wall at (\emph{a}) $f^* = 1$, (\emph{b}) $f^* = 0.1$ and (\emph{c}) $f^* = 0.01$, for $Ra = 10^7$ and $Re_b \approx 5623$.
The dashed lines represent long-time-averaged values of $Nu$ at various modulation frequencies.}
	\label{fig:Nu_bottom}
\end{figure}

Heat accumulates within the boundary layer during the heating phase and is lost during the cooling phase, making temperature profiles dependent on the accumulated heating or cooling. 
In figure \ref{fig:Nu_bottom}, we plot the phase-averaged dimensionless heat flux, represented by the Nusselt number at the bottom wall as $Nu_{\text{bottom}}=-\langle \partial T^* / \partial y^* \rangle_{\text{bottom}, t}$.
Here, $\langle \cdots \rangle_{\text{bottom}, t}$ denotes the ensemble average over the bottom wall and over the time.
At a high frequency of $f^* = 1$ (see figure \ref{fig:Nu_bottom}\emph{a}), the phase-averaged $Nu_{\text{bottom}}$ exhibits substantial oscillation amplitude. 
This occurs because high modulation frequency causes rapid changes in thermal conditions, temporarily enhancing convective heat transfer. 
However, significant peaks and valleys may require robust control mechanisms to manage these rapid changes without causing fatigue due to thermal stress, thereby adding complexity to the system device components in applied thermal engineering applications. 
As the modulation frequency decreases (see figures \ref{fig:Nu_bottom}\emph{b} and \ref{fig:Nu_bottom}\emph{c}), the oscillation amplitude of $Nu_{\text{bottom}}$ decreases, indicating more stable heat transfer with smaller deviations from the baseline. 
This is ideal for applications requiring consistent thermal management without significant fluctuations. 
On the other hand, the long-time-averaged values of Nusselt numbers generally increase as the modulation frequency decreases: 
$Nu$ is 11.58 at $f^* = 1$ while $Nu$ is 20.36 at $f^* = 0.01$ (marked by the dashed lines in figure \ref{fig:Nu_bottom}). 
These results suggest that the average $Nu$ might not fully capture the extreme fluctuations, which are critical for predicting and optimizing heat transfer in various engineering applications. 
 
In addition, we observe the asymmetry in the Nusselt number $Nu_{\text{bottom}}$ during the heating and cooling phases at low modulation frequencies. 
At low frequencies (e.g. $f^* = 0.01$), the system experiences extended periods of heating and cooling, allowing distinct thermal and flow structures to develop in each phase. 
The fluid's thermal inertia causes a delayed response in heat transfer: while the heating phase rapidly amplifies convection, the cooling phase is moderated by the residual thermal energy in the fluid. 
Specifically, during the heating phase, as the bottom wall temperature increases and buoyancy forces enhance, the upward convective currents accelerate rapidly. 
The increasing temperature gradient thins the thermal boundary layer, reducing thermal resistance and enhancing heat transfer efficiency. 
As a result, $Nu_{\text{bottom}}$ rises quickly, forming higher but narrower peaks above the average value. 
During the cooling phase, as the bottom wall temperature decreases and buoyancy forces weaken, the established convective motions persist due to the fluid’s inertia. 
The decreasing temperature gradient thickens the thermal boundary layer, increasing thermal resistance and slowing down heat transfer. 
This results in a more gradual reduction in $Nu_{\text{bottom}}$, producing lower but wider regions below the average value.

Finally, we differentiate between the oscillatory flow induced by wall temperature modulation and the fluctuating fields. 
To that end, we adopt the phase decomposition method, which has been widely applied in the investigation of turbulent flows subjected to periodic forcing, such as channel turbulence or turbulent boundary layers with oscillating walls \citep{choi2002near,ricco2012changes,ebadi2019mean}, oscillatory or pulsating pipe flows \citep{manna2015pulsating,jelly2020direct} and the oscillatory thermal turbulence \citep{wu2021phase,wu2022massive}. 
The key idea of the phase decomposition method is to split the instantaneous field $\mathbb{F}{\left(\mathbf{x},t\right)}$ into the long-time-averaged quantity $\overline{\mathbb{F}}(\mathbf{x})$, the oscillatory quantity $ \tilde{\mathbb{F}}{\left(\mathbf{x},t_n\right)}$  and the turbulent fluctuation $\mathbb{F}^{\prime}(\mathbf{x},t)$, which is written as 
\begin{equation}
\mathbb{F}(\mathbf{x},t)=\overline{\mathbb{F}}(\mathbf{x})+\overset{\sim}{\operatorname*{\mathbb{F}}}(\mathbf{x},t_n)+\mathbb{F}^{\prime}(\mathbf{x},t)
\end{equation}
Here, the oscillatory quantity is calculated as $\widetilde{\mathbb{F}}(\mathbf{x},t_n)=\left\langle\mathbb{F}(\mathbf{x})\right\rangle_\phi-\overline{\mathbb{F}}(\mathbf{x})$. 
We divide the oscillating period $T_{\text{period}}$ into $M$ evenly spaced intervals, and $t_n=(n-1)T_{\text{period}}/M$ is the time corresponding to the phase angle of $\phi_n=2(n-1)\pi/M$. 
Previously, \citet{yang2020periodically} investigated the thermal Stokes problem in pure RB convection, where the temperature oscillated in modulated pure RB convection. 
The oscillatory temperature can be obtained analytically as $ \tilde{T}(y,t_n)=Ae^{-y^*/\lambda_s^*}\sin(\phi_n-y^*/\lambda_s^*)$, with the Stokes thermal boundary layer thickness being $\lambda_s^*=\pi^{-1/2}f^{*-1/2}Ra^{-1/4}Pr^{-1/4}$. 
Here, the term $Ae^{-y^*/\lambda_s^*}$ indicates that the effect of the oscillating boundary temperature decays exponentially with distance from the boundary, while the term $\sin(\phi_n-y^*/\lambda_s^*)$ indicates a phase lag of $y^*/\lambda_s^*$ in the temperature oscillation as we move away from the boundary.

\begin{figure}
	\centerline{\includegraphics[width=0.8\textwidth]{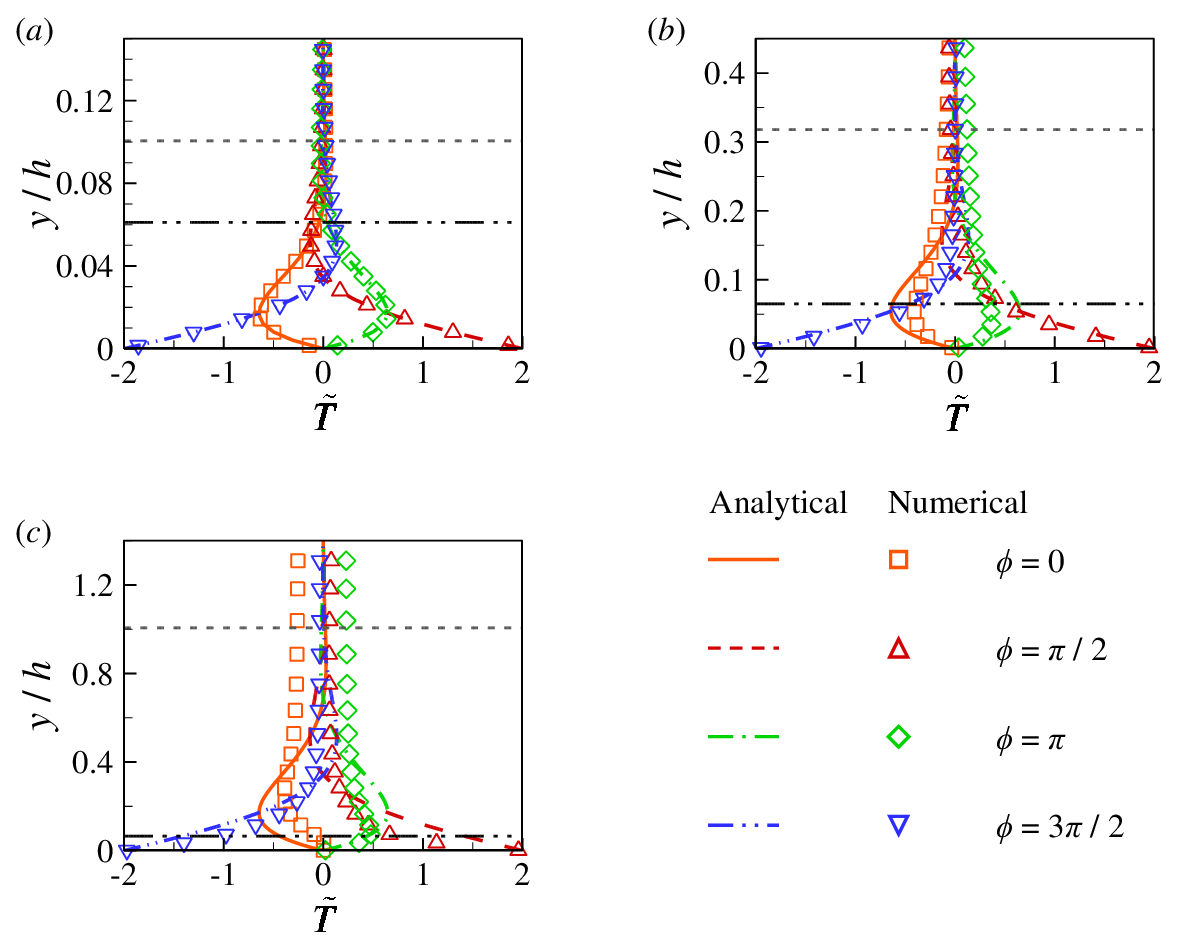}}
	\caption{Comparison of oscillatory temperature from analytical and numerical solutions, at (\emph{a}) $f^* = 1$, (\emph{b}) $f^* = 0.1$ and (\emph{c}) $f^* = 0.01$,  for $Ra = 10^7$ and $Re_b \approx 5623$.
The grey dashed lines show the depth where the oscillating temperature's amplitude drops to 1\% of its maximum value, and the black dash-dotted lines show the thermal boundary layer thickness.
}
	\label{fig:oscillatoryT}
\end{figure}

We compare the analytical and numerical solutions of oscillatory temperature in mixed PRB convection, as shown in figure \ref{fig:oscillatoryT}. 
A good agreement is observed at a high frequency of $f^* = 1$ (see figure \ref{fig:oscillatoryT}\emph{a}); 
however, deviations occur at lower frequencies of $f^* = 0.1$ and 0.01 (see figures \ref{fig:oscillatoryT}\emph{b} and \ref{fig:oscillatoryT}\emph{c}). 
These discrepancies may be attributed to two factors.
First, the penetration depth of the temperature disturbance induced by the oscillating wall temperature is inversely proportional to the modulation frequency. 
Consequently, at lower frequencies, the temperature disturbance penetrates deeper into the flow, potentially exceeding the thermal boundary layer thickness. 
Within the thermal boundary, we can assume a heat conduction profile for the temperature, which aligns with the analytical model \citep{yang2020periodically}. 
Outside the thermal boundary, the effects of convection become significant, causing deviations from the analytical predictions. 
In figure \ref{fig:oscillatoryT}, we mark the depth where the oscillating temperature's amplitude drops to 1\% of its maximum value (approximately equal to 4.6$\lambda_s^*$), as well as the thermal boundary layer thickness (determined from the peak positions of the r.m.s. temperature profile). 
The results show that at a higher frequency of $f^* = 1$, the oscillating temperature penetration depth is of the same order of magnitude as the thermal boundary layer thickness.
However, at a lower frequency of $f^* = 0.01$, the penetration depth is an order of magnitude thicker than the thermal boundary layer thickness.

Moreover, our simulations included an imposed pressure gradient (achieved via an equivalent body force), which introduces mean shear that affects the velocity field throughout the fluid domain, including within the thermal boundary layer. 
This shear enhances convective transport parallel to the wall, modifying the temperature profile. 
A critical factor here is the ratio of the thermal diffusion timescale to the convective timescale induced by the mean shear. 
We denote the thermal diffusion timescale $t_{\text{diff}}$ as the characteristics time it takes for heat to diffuse across the penetration depth $\lambda_s$, with $t_{\text{diff}} \propto \lambda_s^2 / \alpha$ (where $\alpha$ is the thermal diffusivity). 
We also denote the convective timescale $t_{\text{conv}}$ as the characteristics time it takes for fluid particles to be advected by the mean shear across the penetration depth, with $t_{\text{conv}} \propto \lambda_s / U$ (where $U$ is the characteristic velocity of the Poiseuille flow near the wall). 
Comparing these timescales, we have the ratio $t_{\text{diff}} / t_{\text{conv}}\propto f^{*-1/2}$. 
As the modulation frequency $f^*$ decreases, the thermal diffusion timescale $t_{\text{diff}}$ becomes larger compared with the convective timescale $t_{\text{conv}}$, indicating that convection becomes more significant within the penetration depth. 
This increased significance of convection, driven by the mean shear, contributes to the discrepancies between the analytical and simulation results.

\section {Conclusion}
\label{sec:4 Conclusion}
In this work, we have performed DNS of mixed convection in turbulent PRB channels with sinusoidal wall temperature modulation. 
High-frequency wall temperature oscillations resulted in a relatively unchanged flow structure, while low-frequency oscillations caused the flow structures to adapt over time, forming stable stratified layers during the cooling phase. 
Using the POD method, we identified the most energetic mode as a dominant streamwise unidirectional shear flow. 
Regardless of modulation frequency, streamwise-oriented large-scale rolls appeared as higher POD modes. 
Streamwise-oriented rolls were strongly correlated with wall temperature variations at lower frequencies, indicating a significant influence on roll strength. 
We also tracked the movements of roll centres, showing that large-scale rolls exhibit non-stationary behaviour influenced by the periodic boundary conditions of the computational domain. 
In addition, vertical convective heat flux analysis revealed counter-gradient heat transfer driven by thermal plumes at high frequencies and by bulk roll dynamics at low frequencies.

We then explored the impact of wall temperature modulation on long-time-averaged statistics of mean flow and temperature. 
The friction coefficient showed less than a 15\% variation with modulation frequency.
In contrast, the Nusselt number increased as the frequency decreased, particularly at higher Rayleigh numbers, with an increase up to 96\%. 
High-frequency modulation minimally affected the mean profiles of streamwise velocity and temperature.
However, lower frequencies increased the velocity near the top wall and decreased it near the bottom wall, and the temperature profile deviated from that of canonical RB convection. 
Total shear stress varied non-linearly at high frequencies but linearly at lower frequencies, resembling the behaviour of a pure turbulent channel without convection. 
This difference arises because high-frequency modulation enhances convective effects near the wall, while lower frequencies reduce these effects by forming a stable stratified layer.

Finally, we analysed phase-averaged statistics to understand variability during different phases of the flow cycle. 
At high modulation frequency, the phase-averaged streamwise velocity profile remained stable.
However, at low frequencies, the velocity decreased near the wall and increased away from it due to weakened turbulence. 
The TKE profiles were consistently high near the wall at high frequency, but they were lower during the cooling phases at low frequency. 
The TKE budget analysis revealed that shear dominated TKE production near the wall during heating, while buoyancy dominated in the bulk region; both TKE productions were nearly zero during cooling. 
High modulation frequency confined temperature deviations to the region near the wall, whereas lower frequency affected temperatures both near the wall and in the bulk region.
During heating, temperature decreased upward in the boundary layer, leading to unstable stratification. 
During cooling, temperature increased upward, creating a stable boundary layer that suppressed turbulence and formed a residual layer. 
Phase-averaged Nusselt numbers showed substantial oscillation amplitude at high frequency due to rapid thermal changes, while oscillations decreased with lower frequency, leading to more stable and efficient heat transfer.

Overall, our study highlights the complex interplay between wall temperature modulation frequency and flow dynamics in mixed convection, providing insights relevant to meteorology and heat transfer engineering applications.
However, we acknowledge that the limited spanwise domain size may inhibit the development of the largest coherent structures and alter the flow morphology compared with larger domains representative of atmospheric flows. 
Furthermore, the limited range of flow parameters restricts our ability to fully mimic the extreme flow conditions in atmospheric currents. 
Therefore, caution should be exercised when generalizing these results to larger-scale flows, such as those in the atmospheric boundary layer. 
Future studies with larger computational domains and higher Rayleigh and Reynolds numbers could provide additional insights into scale-dependent behaviours and further validate the applicability of our conclusions to real-world atmospheric flows.

\backsection[Supplementary movies]{\label{SupMat}A supplementary movies are available at \url{https://doi.org/10.1017/jfm.2025.22}.}

\backsection[Funding]{This work was supported by the National Natural Science Foundation of China (NSFC) through grants nos. 12272311, 12125204, 12388101; the Young Elite Scientists Sponsorship Program by CAST (2023QNRC001); and the 111 project of China (project no. B17037).}

\backsection[Declaration of interests]{The authors report no conﬂict of interest.}


\backsection[Author ORCIDs]{Ao Xu, https://orcid.org/0000-0003-0648-2701;	Heng-Dong Xi, https://orcid.org/0000-0002-2999-2694}


\appendix

\section{Cross-validation of solvers against benchmark data}\label{appA}
\begin{figure}
	\centerline{\includegraphics[width=0.95\textwidth]{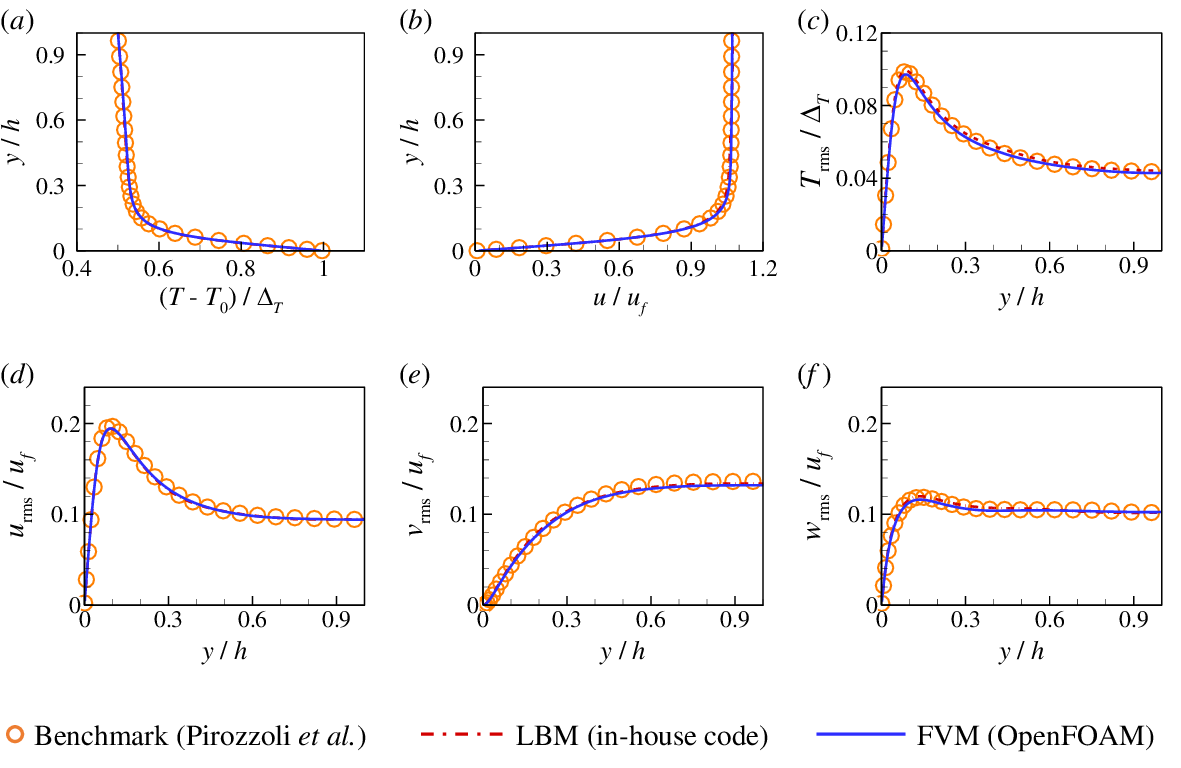}}
	\caption{Comparison of data from \citet{pirozzoli2017mixed} as benchmark, our in-house code based on LBM, and OpenFOAM code based on FVM. 
Mean profiles of (\emph{a}) temperature and (\emph{b}) streamwise velocity, mean-square fluctuation of (\emph{c}) temperature, (\emph{d}) streamwise velocity, (\emph{e}) wall-normal velocity and (\emph{f}) spanwise velocity along the bottom half-height of the channel for $Re_{b} \approx 3162$, $Ra = 10^{7}$ and $Pr = 1$.
}
	\label{fig:validation}
\end{figure}

Here, we validate the consistency of results among the benchmark data from \citet{pirozzoli2017mixed}, our in-house solver based on the LBM, and the OpenFOAM solver based on the FVM.
As shown in figure \ref{fig:validation}, we compare flow quantities including the mean profiles of temperature and streamwise velocity, as well as the r.m.s. fluctuations of temperature, streamwise velocity, wall-normal velocity and spanwise velocity along the bottom half-height of the channel. 
The results demonstrate good agreement between all three datasets, confirming the accuracy and reliability of our simulation results.
Furthermore, table \ref{tab:validation} provides a detailed quantitative comparison of flow quantities. 
The numerical comparison shows minimal deviations, with discrepancies remaining within acceptable ranges for turbulent simulations, further reinforcing the consistency of the results.

\begin{table}
	\begin{center}
		\def~{\hphantom{0}}
		\begin{tabular}{cccccccc}
			Flow database                           &  $Re_\tau$  &  $Nu$  &  $C_{f}$  &  $N_x$  &  $N_y$  &  $N_z$  \\
            Benchmark (Pirozzoli \emph{et al.})  &  142.59  &  11.622  &  $1.62\times 10^{-2}$  & 256  &  256  &  128 \\
            LBM (in-house code)                     &  142.08  &  11.713  &  $1.62\times 10^{-2}$  & 420  &  210  &  210 \\
            FVM (OpenFOAM)                          &  141.62  &  11.616  &  $1.60\times 10^{-2}$  & 256  &  256  &  128 \\
		\end{tabular}%
		\caption{Quantitative comparison of flow quantities among benchmark, LBM and FVM solvers. 
The columns from left to right indicate the following: flow database; friction Reynolds number ($Re_{\tau}$); Nusselt number ($Nu$); skin friction coefficient ($C_{f}$); grid number in the streamwise, wall-normal, and spanwise directions ($N_{x}$, $N_{y}$, $N_{z}$).}
		\label{tab:validation}
	\end{center}
\end{table}

\section{Parameter estimation for desert atmospheric currents}\label{appB}
We consider typical desert atmospheric conditions, where the characteristic height $H$ is approximately 1000 m, representing the atmospheric boundary layer height. 
For air at a mean temperature $T_{\text{mean}} = 300$ K, the thermal expansion coefficient is $\beta = 3.36 \times 10^{-3}$ K$^{-1}$, the kinematic viscosity is $\nu =1.57 \times 10^{-5}$ m$^{2}$/s and the thermal diffusivity is $\alpha = 2.22 \times 10^{-5}$ m$^{2}$/s. 
The temperature difference across the atmospheric boundary layer can be $\Delta_T \approx 20$ K or more.
The gravitational acceleration is $g = 9.8$ m/s$^{2}$.

i) Modulation amplitude. 
In desert regions, surface temperatures fluctuate significantly due to strong solar heating during the day and rapid radiative cooling at night, leading to both unstable and stable stratification over 24 hours. 
To mimic this, we vary the bottom wall temperature as $T_{\text{bottom}} = T_{\text{hot}} + A (T_{\text{hot}} - T_{\text{cold}}) \sin(2\pi ft)$, representing a sinusoidal variation around $T_{\text{hot}} = [(T_{\text{bottom}})_{\min} + (T_{\text{bottom}})_{\max}] / 2$. 
Data from the moderate-resolution imaging spectroradiometer observations \citep{sharifnezhadazizi2019global} indicate that land surface temperatures in the Sahara Desert range between $(T_{\text{bottom}})_{\min} = 280$ K and $(T_{\text{bottom}})_{\max} = 310$ K, suggesting an average bottom wall temperature of $T_{\text{hot}} = 295$ K. 
Above the atmospheric boundary layer, diurnal cycles are absent, so the top wall temperature is fixed as $T_{\text{top}} = T_{\text{cold}}$.  
Assuming a lapse rate-adjusted temperature at the top boundary, we estimate $T_{\text{top}} = T_{\text{cold}} \approx T_{\text{hot}} – 6.5$ K/km $\times$ 1 km $\approx 288.5$ K. 
Thus, $A = \left[ (T_{\text{bottom}})_{\max} - T_{\text{hot}} \right] / (T_{\text{hot}} - T_{\text{cold}}) \approx 2.3$, closely matching our chosen value of $A=2$.

ii) Modulation frequency. 
The diurnal frequency associated with the day-night cycle is relatively low. 
On Earth, a 24-hour diurnal period corresponds to a frequency of approximately $f_{\text{diurnal}} = 1 / (24 \times 3600 \ \text{s}) \approx 1.16 \times 10^{-5}$ Hz. 
To relate this to our dimensionless frequency $f^*$, we estimate the free-fall time scale as $t_f=\sqrt{H/(g \beta \Delta_T)} \approx 39$ s, yielding a dimensionless diurnal frequency of $f^* = f_{\text{diurnal}} t_{f} \approx 4.5 \times 10^{-4}$. 
This value is two orders of magnitude smaller than the lower limit of our study (i.e. $f^* = 0.01$). 
Exploring lower frequencies would require significantly more computational resources to ensure the convergence of phase-averaged statistics over extended simulation times.

iii) Prandlt number. 
The Prandlt number characterizes the thermophysical properties of a fluid and is defined as $Pr=\nu/\alpha$. 
It quantifies the ratio of viscous diffusion to thermal diffusivity. 
For air, we have $Pr \approx 0.71$, which closely matches our settings.

iv) Rayleigh number. 
The Rayleigh number characterizes the intensity of convection in a fluid system and is defined as $Ra=g\beta \Delta_T H^3/(\nu \alpha)$. 
It quantifies the balance between buoyancy-driven forces and diffusive transport (viscous and thermal diffusion). 
For typical desert atmospheric conditions, we have $Ra \approx 2 \times 10^{18}$.

v) Friction Reynolds number. 
The friction Reynolds number characterizes turbulent shear near the wall and is defined as $Re_\tau=u_\tau H/ \nu$. 
To estimate the friction velocity $u_\tau$, we use the logarithmic wind profile $U(z)=u_\tau / \kappa \ln (z/z_0)$, where $\kappa \approx 0.4$ is the von Kármán constant and $z_0$  is the roughness length. 
For desert surfaces, we estimate $z_0=0.01$ m. 
During sandstorms, wind speeds reach 13.9 m/s or higher (i.e. Level 7 on the Beaufort scale, measured at $z = 10$ m), resulting in a friction velocity of $u_\tau \approx 0.8$ m/s, giving $Re_\tau \approx  5 \times 10^7$.

vi) Bulk Reynolds number. 
The bulk Reynolds number represents the ratio of inertial forces to viscous forces in the flow, considering the average motion of the fluid across the entire flow domain, and is defined as $Re_b=u_b H/\nu$. 
Using the logarithmic wind profile, we obtain the mean bulk velocity $u_b$ as  $\left(\int_{z_0}^{H}U(z)dz \right)/H \approx u_{\tau}/\kappa\left[\ln(H)-1-\ln(z_{0})\right] \approx 21$ m/s, leading to $Re_b \approx 1 \times 10^9$. 
The corresponding bulk Richardson number is $Ri_b=Ra/(Re_b^2 Pr) \sim O(1)$, suggesting that global buoyancy and shear effects are comparable.
 
These high Rayleigh and Reynolds numbers align with the expected intense turbulence and large-scale atmospheric motions, indicating an extreme flow condition. 
There remains a significant gap between current computational capabilities and the requirements for DNS or large-eddy simulation of sheared thermal turbulence under such extreme parameters.


\begin{thebibliography}{78}
\expandafter\ifx\csname natexlab\endcsname\relax\def\natexlab#1{#1}\fi
\def\au#1{#1} \def\ed#1{#1} \def\yr#1{#1}\def\at#1{#1}\def\jt#1{\textit{#1}}
  \def\bt#1{#1}\def\bvol#1{\textbf{#1}} \def\vol#1{#1} \def\pg#1{#1}
  \def\publ#1{#1}\def\arxiv#1{#1}\def\org#1{#1}\def\st#1{\textit{#1}}

\bibitem[Ahlers {\em et~al.\/}(2009)Ahlers, Grossmann \& Lohse]{ahlers2009heat}
{\sc \au{Ahlers, G.}, \au{Grossmann, S.} \& \au{Lohse, D.}} \yr{2009}
  \at{{Heat transfer and large scale dynamics in turbulent Rayleigh-B{\'e}}nard
  convection}.  \jt{Rev. Mod. Phys.}  \bvol{81}~(2),  \pg{503--537}.

\bibitem[Andreotti {\em et~al.\/}(2009)Andreotti, Fourriere, Ould-Kaddour,
  Murray \& Claudin]{andreotti2009giant}
{\sc \au{Andreotti, B.}, \au{Fourriere, A.}, \au{Ould-Kaddour, F.}, \au{Murray,
  B.} \& \au{Claudin, P.}} \yr{2009}  \at{{Giant aeolian dune size determined
  by the average depth of the atmospheric boundary layer}}.  \jt{Nature}
  \bvol{457}~(7233),  \pg{1120--1123}.

\bibitem[Berkooz {\em et~al.\/}(1993)Berkooz, Holmes \&
  Lumley]{berkooz1993proper}
{\sc \au{Berkooz, G.}, \au{Holmes, P.} \& \au{Lumley, J.~L.}} \yr{1993}
  \at{{The proper orthogonal decomposition in the analysis of turbulent
  flows}}.  \jt{Annu. Rev. Fluid Mech.}  \bvol{25}~(1),  \pg{539--575}.

\bibitem[Bernardini {\em et~al.\/}(2014)Bernardini, Pirozzoli \&
  Orlandi]{bernardini2014velocity}
{\sc \au{Bernardini, M.}, \au{Pirozzoli, S.} \& \au{Orlandi, P.}} \yr{2014}
  \at{{Velocity statistics in turbulent channel flow up to $Re_{\tau}=4000$}}.
  \jt{J. Fluid Mech.}  \bvol{742},  \pg{171--191}.

\bibitem[Blass {\em et~al.\/}(2021)Blass, Tabak, Verzicco, Stevens \&
  Lohse]{blass2021effect}
{\sc \au{Blass, A.}, \au{Tabak, P.}, \au{Verzicco, R.}, \au{Stevens, R.J.A.M.}
  \& \au{Lohse, D.}} \yr{2021}  \at{{The effect of Prandtl number on turbulent
  sheared thermal convection}}.  \jt{J. Fluid Mech.}  \bvol{910},  \pg{A37}.

\bibitem[Blass {\em et~al.\/}(2020)Blass, Zhu, Verzicco, Lohse \&
  Stevens]{blass2020flow}
{\sc \au{Blass, A.}, \au{Zhu, X.}, \au{Verzicco, R.}, \au{Lohse, D.} \&
  \au{Stevens, R.J.A.M.}} \yr{2020}  \at{{Flow organization and heat transfer
  in turbulent wall sheared thermal convection}}.  \jt{J. Fluid Mech.}
  \bvol{897},  \pg{A22}.

\bibitem[Brown(1980)]{brown1980longitudinal}
{\sc \au{Brown, R.~A.}} \yr{1980}  \at{{Longitudinal instabilities and
  secondary flows in the planetary boundary layer: A review}}.  \jt{Rev.
  Geophys.}  \bvol{18}~(3),  \pg{683--697}.

\bibitem[Castillo-Castellanos {\em et~al.\/}(2019)Castillo-Castellanos,
  Sergent, Podvin \& Rossi]{castillo2019cessation}
{\sc \au{Castillo-Castellanos, A.}, \au{Sergent, A.}, \au{Podvin, B.} \&
  \au{Rossi, M.}} \yr{2019}  \at{{Cessation and reversals of large-scale
  structures in square Rayleigh--B{\'e}}nard cells}.  \jt{J. Fluid Mech.}
  \bvol{877},  \pg{922--954}.

\bibitem[Caulfield(2021)]{caulfield2021layering}
{\sc \au{Caulfield, C.~P.}} \yr{2021}  \at{{Layering, instabilities, and mixing
  in turbulent stratified flows}}.  \jt{Annu. Rev. Fluid Mech.}  \bvol{53}~(1),
   \pg{113--145}.

\bibitem[Chen \& Sreenivasan(2023)]{chen2022reynolds}
{\sc \au{Chen, X.} \& \au{Sreenivasan, K.R.}} \yr{2023}  \at{{Reynolds number
  asymptotics of wall-turbulence fluctuations.}}  \jt{J. Fluid Mech.}
  \bvol{976},  \pg{A21}.

\bibitem[Chilla {\em et~al.\/}(2012)Chilla, Evrard \&
  Schulz]{chilla2012territoriality}
{\sc \au{Chilla, T.}, \au{Evrard, E.} \& \au{Schulz, C.}} \yr{2012}  \at{{On
  the territoriality of cross-border cooperation: “Institutional Mapping”
  in a multi-level context}}.  \jt{Eur. Plan. Stud.}  \bvol{20}~(6),
  \pg{961--980}.

\bibitem[Choi(2002)]{choi2002near}
{\sc \au{Choi, K.-S.}} \yr{2002}  \at{{Near-wall structure of turbulent
  boundary layer with spanwise-wall oscillation}}.  \jt{Phys. Fluids}
  \bvol{14}~(7),  \pg{2530--2542}.

\bibitem[Cossu(2022)]{cossu2022onset}
{\sc \au{Cossu, C.}} \yr{2022}  \at{{Onset of large-scale convection in
  wall-bounded turbulent shear flows}}.  \jt{J. Fluid Mech.}  \bvol{945},
  \pg{A33}.

\bibitem[Dror {\em et~al.\/}(2023)Dror, Koren, Liu \&
  Altaratz]{dror2023convective}
{\sc \au{Dror, T.}, \au{Koren, I.}, \au{Liu, H.} \& \au{Altaratz, O.}}
  \yr{2023}  \at{{Convective steady state in shallow cloud fields}}.  \jt{Phys.
  Rev. Lett.}  \bvol{131}~(13),  \pg{134201}.

\bibitem[Dupont \& Patton(2022)]{dupont2022influence}
{\sc \au{Dupont, S.} \& \au{Patton, E.~G.}} \yr{2022}  \at{{On the influence of
  large-scale atmospheric motions on near-surface turbulence: Comparison
  between flows over low-roughness and tall vegetation canopies}}.
  \jt{Bound.-Layer Meteor.}  \bvol{184}~(2),  \pg{195--230}.

\bibitem[Ebadi {\em et~al.\/}(2019)Ebadi, White, Pond \& Dubief]{ebadi2019mean}
{\sc \au{Ebadi, A.}, \au{White, C.~M.}, \au{Pond, I.} \& \au{Dubief, Y.}}
  \yr{2019}  \at{{Mean dynamics and transition to turbulence in oscillatory
  channel flow}}.  \jt{J. Fluid Mech.}  \bvol{880},  \pg{864--889}.

\bibitem[Garai {\em et~al.\/}(2014)Garai, Kleissl \& Sarkar]{garai2014flow}
{\sc \au{Garai, A.}, \au{Kleissl, J.} \& \au{Sarkar, S.}} \yr{2014}  \at{{Flow
  and heat transfer in convectively unstable turbulent channel flow with
  solid-wall heat conduction}}.  \jt{J. Fluid Mech.}  \bvol{757},  \pg{57--81}.

\bibitem[Gasteuil {\em et~al.\/}(2007)Gasteuil, Shew, Gibert, Chill{\`a},
  Castaing \& Pinton]{gasteuil2007lagrangian}
{\sc \au{Gasteuil, Y.}, \au{Shew, W.~L.}, \au{Gibert, M.}, \au{Chill{\`a}, F.},
  \au{Castaing, B.} \& \au{Pinton, J.-F.}} \yr{2007}  \at{{Lagrangian
  temperature, velocity, and local heat flux measurement in
  Rayleigh-B{\'e}}nard convection}.  \jt{Phys. Rev. Lett.}  \bvol{99}~(23),
  \pg{234302}.

\bibitem[Graham \& Floryan(2021)]{graham2021exact}
{\sc \au{Graham, M.~D.} \& \au{Floryan, D.}} \yr{2021}  \at{{Exact coherent
  states and the nonlinear dynamics of wall-bounded turbulent flows}}.
  \jt{Annu. Rev. Fluid Mech.}  \bvol{53}~(1),  \pg{227--253}.

\bibitem[Hanna(1969)]{hanna1969formation}
{\sc \au{Hanna, S.~R.}} \yr{1969}  \at{{The formation of longitudinal sand
  dunes by large helical eddies in the atmosphere}}.  \jt{J. Appl. Meteorol.
  Climatol.}  \bvol{8}~(6),  \pg{874--883}.

\bibitem[Howland {\em et~al.\/}(2024)Howland, Yerragolam, Verzicco \&
  Lohse]{howland2024turbulent}
{\sc \au{Howland, C.J.}, \au{Yerragolam, G.S.}, \au{Verzicco, R.} \& \au{Lohse,
  D.}} \yr{2024}  \at{Turbulent mixed convection in vertical and horizontal
  channels}.  \jt{J. Fluid Mech.}  \bvol{998},  \pg{A48}.

\bibitem[Huang \& Zhou(2013)]{huang2013counter}
{\sc \au{Huang, Y.-X.} \& \au{Zhou, Q.}} \yr{2013}  \at{{Counter-gradient heat
  transport in two-dimensional turbulent Rayleigh--B{\'e}}nard convection}.
  \jt{J. Fluid Mech.}  \bvol{737},  \pg{R3}.

\bibitem[Hunt {\em et~al.\/}(1988)Hunt, Wray \& Moin]{hunt1988eddies}
{\sc \au{Hunt, J. C.~R.}, \au{Wray, A.~A.} \& \au{Moin, P.}} \yr{1988} {Eddies,
  streams, and convergence zones in turbulent flows}.  \bt{In {\em Studying
  Turbulence Using Numerical Simulation Databases, 2. Proceedings of the 1988
  Summer Program\/}},  \pg{pp. 193--208}.

\bibitem[Iida \& Kasagi(1997)]{iida1997direct}
{\sc \au{Iida, O.} \& \au{Kasagi, N.}} \yr{1997}  \at{{Direct numerical
  simulation of unstably stratified turbulent channel flow}}.  \jt{ASME J. Heat
  Transf.}  \bvol{119}~(1),  \pg{53--61}.

\bibitem[Jayaraman \& Brasseur(2021)]{jayaraman2021transition}
{\sc \au{Jayaraman, B.} \& \au{Brasseur, J.~G.}} \yr{2021}  \at{{Transition in
  atmospheric boundary layer turbulence structure from neutral to convective,
  and large-scale rolls}}.  \jt{J. Fluid Mech.}  \bvol{913},  \pg{A42}.

\bibitem[Jelly {\em et~al.\/}(2020)Jelly, Chin, Illingworth, Monty, Marusic \&
  Ooi]{jelly2020direct}
{\sc \au{Jelly, T.~O.}, \au{Chin, R.~C.}, \au{Illingworth, S.~J.}, \au{Monty,
  J.~P.}, \au{Marusic, I.} \& \au{Ooi, A.}} \yr{2020}  \at{{A direct comparison
  of pulsatile and non-pulsatile rough-wall turbulent pipe flow}}.  \jt{J.
  Fluid Mech.}  \bvol{895},  \pg{R3}.

\bibitem[Jiang {\em et~al.\/}(2020)Jiang, Zhu, Wang, Huisman \&
  Sun]{jiang2020supergravitational}
{\sc \au{Jiang, H.}, \au{Zhu, X.}, \au{Wang, D.}, \au{Huisman, S.G.} \&
  \au{Sun, C.}} \yr{2020}  \at{{Supergravitational turbulent thermal
  convection}}.  \jt{Sci. Adv.}  \bvol{6}~(40),  \pg{eabb8676}.

\bibitem[Jim{\'e}nez(2012)]{jimenez2012cascades}
{\sc \au{Jim{\'e}nez, J.}} \yr{2012}  \at{{Cascades in wall-bounded
  turbulence}}.  \jt{Annu. Rev. Fluid Mech.}  \bvol{44}~(1),  \pg{27--45}.

\bibitem[Jim{\'e}nez(2013)]{jimenez2013near}
{\sc \au{Jim{\'e}nez, J.}} \yr{2013}  \at{{Near-wall turbulence}}.  \jt{Phys.
  Fluids}  \bvol{25}~(10),  \pg{101302}.

\bibitem[Jin \& Xia(2008)]{jin2008experimental}
{\sc \au{Jin, X.-L.} \& \au{Xia, K.-Q.}} \yr{2008}  \at{{An experimental study
  of kicked thermal turbulence}}.  \jt{J. Fluid Mech.}  \bvol{606},
  \pg{133--151}.

\bibitem[Kays(1994)]{kays1994turbulent}
{\sc \au{Kays, W.~M.}} \yr{1994}  \at{{Turbulent Prandtl number. Where are
  we?}}  \jt{ASME J. Heat Transf.}  \bvol{116}~(2),  \pg{284--295}.

\bibitem[Kok {\em et~al.\/}(2012)Kok, Parteli, Michaels \&
  Karam]{kok2012physics}
{\sc \au{Kok, J.~F.}, \au{Parteli, E. J.~R.}, \au{Michaels, T.~I.} \&
  \au{Karam, D.~B.}} \yr{2012}  \at{{The physics of wind-blown sand and dust}}.
   \jt{Rep. Prog. Phys.}  \bvol{75}~(10),  \pg{106901}.

\bibitem[Komen {\em et~al.\/}(2014)Komen, Shams, Camilo \&
  Koren]{komen2014quasi}
{\sc \au{Komen, E.}, \au{Shams, A.}, \au{Camilo, L.} \& \au{Koren, B.}}
  \yr{2014}  \at{{Quasi-DNS capabilities of OpenFOAM for different mesh
  types}}.  \jt{Comput. Fluids}  \bvol{96},  \pg{87--104}.

\bibitem[Komen {\em et~al.\/}(2017)Komen, Camilo, Shams, Geurts \&
  Koren]{komen2017quantification}
{\sc \au{Komen, E. M.~J.}, \au{Camilo, L.~H.}, \au{Shams, A.}, \au{Geurts,
  B.~J.} \& \au{Koren, B.}} \yr{2017}  \at{{A quantification method for
  numerical dissipation in quasi-DNS and under-resolved DNS, and effects of
  numerical dissipation in quasi-DNS and under-resolved DNS of turbulent
  channel flows}}.  \jt{J. Comput. Phys.}  \bvol{345},  \pg{565--595}.

\bibitem[Komen {\em et~al.\/}(2023)Komen, Mathur, Roelofs, Merzari \&
  Tiselj]{komen2023status}
{\sc \au{Komen, E. M.~J.}, \au{Mathur, A.}, \au{Roelofs, F.}, \au{Merzari, E.}
  \& \au{Tiselj, I.}} \yr{2023}  \at{{Status, perspectives, and added value of
  high fidelity simulations for safety and design}}.  \jt{Nucl. Eng. Des.}
  \bvol{401},  \pg{112082}.

\bibitem[Komori {\em et~al.\/}(1982)Komori, Ueda, Ogino \&
  Mizushina]{komori1982turbulence}
{\sc \au{Komori, S.}, \au{Ueda, H.}, \au{Ogino, F.} \& \au{Mizushina, T.}}
  \yr{1982}  \at{{Turbulence structure in unstably-stratified open-channel
  flow}}.  \jt{Phys. Fluids}  \bvol{25}~(9),  \pg{1539--1546}.

\bibitem[Kooij {\em et~al.\/}(2018)Kooij, Botchev, Frederix, Geurts, Horn,
  Lohse, van~der Poel, Shishkina, Stevens \& Verzicco]{kooij2018comparison}
{\sc \au{Kooij, G.~L.}, \au{Botchev, M.~A.}, \au{Frederix, E. M.~A.},
  \au{Geurts, B.~J.}, \au{Horn, S.}, \au{Lohse, D.}, \au{van~der Poel, E.~P.},
  \au{Shishkina, O.}, \au{Stevens, R.J.A.M.} \& \au{Verzicco, R.}} \yr{2018}
  \at{{Comparison of computational codes for direct numerical simulations of
  turbulent Rayleigh--B{\'e}}nard convection}.  \jt{Comput. Fluids}
  \bvol{166},  \pg{1--8}.

\bibitem[Li {\em et~al.\/}(2022)Li, Chen, Xu \& Xi]{li2022counter}
{\sc \au{Li, Y.-Z.}, \au{Chen, X.}, \au{Xu, A.} \& \au{Xi, H.-D.}} \yr{2022}
  \at{{Counter-flow orbiting of the vortex centre in turbulent thermal
  convection}}.  \jt{J. Fluid Mech.}  \bvol{935},  \pg{A19}.

\bibitem[Liu {\em et~al.\/}(2014)Liu, He, Wang \& Zhang]{liu2014advances}
{\sc \au{Liu, H.}, \au{He, M.}, \au{Wang, B.} \& \au{Zhang, Q.}} \yr{2014}
  \at{{Advances in low-level jet research and future prospects}}.  \jt{J.
  Meteorol. Res.}  \bvol{28}~(1),  \pg{57--75}.

\bibitem[Lohse(2024)]{lohse2024asking}
{\sc \au{Lohse, D.}} \yr{2024}  \at{{Asking the right questions on
  Rayleigh–B{\'e}}nard turbulence}.  \jt{J. Fluid Mech.}  \bvol{1000},
  \pg{F3}.

\bibitem[Lohse \& Shishkina(2024)]{lohse2024ultimate}
{\sc \au{Lohse, D.} \& \au{Shishkina, O.}} \yr{2024}  \at{{Ultimate
  Rayleigh-B{\'e}}nard turbulence}.  \jt{Rev. Mod. Phys.}  \bvol{96}~(3),
  \pg{035001}.

\bibitem[Lohse \& Xia(2010)]{lohse2010small}
{\sc \au{Lohse, D.} \& \au{Xia, K.-Q.}} \yr{2010}  \at{{Small-scale properties
  of turbulent Rayleigh-B{\'e}}nard convection}.  \jt{Annu. Rev. Fluid Mech.}
  \bvol{42}~(1),  \pg{335--364}.

\bibitem[Madhusudanan {\em et~al.\/}(2022)Madhusudanan, Illingworth, Marusic \&
  Chung]{madhusudanan2022navier}
{\sc \au{Madhusudanan, A.}, \au{Illingworth, S.~J.}, \au{Marusic, I.} \&
  \au{Chung, D.}} \yr{2022}  \at{{Navier-Stokes--based linear model for
  unstably stratified turbulent channel flows}}.  \jt{Phys. Rev. Fluids}
  \bvol{7}~(4),  \pg{044601}.

\bibitem[Manna {\em et~al.\/}(2015)Manna, Vacca \&
  Verzicco]{manna2015pulsating}
{\sc \au{Manna, M.}, \au{Vacca, A.} \& \au{Verzicco, R.}} \yr{2015}
  \at{{Pulsating pipe flow with large-amplitude oscillations in the very high
  frequency regime. Part 2. Phase-averaged analysis}}.  \jt{J. Fluid Mech.}
  \bvol{766},  \pg{272--296}.

\bibitem[Marusic {\em et~al.\/}(2021)Marusic, Chandran, Rouhi, Fu, Wine,
  Holloway, Chung \& Smits]{marusic2021energy}
{\sc \au{Marusic, I.}, \au{Chandran, D.}, \au{Rouhi, A.}, \au{Fu, M.~K.},
  \au{Wine, D.}, \au{Holloway, B.}, \au{Chung, D.} \& \au{Smits, A.~J.}}
  \yr{2021}  \at{{An energy-efficient pathway to turbulent drag reduction}}.
  \jt{Nat. Commun.}  \bvol{12}~(1),  \pg{5805}.

\bibitem[Marusic {\em et~al.\/}(2010)Marusic, McKeon, Monkewitz, Nagib, Smits
  \& Sreenivasan]{marusic2010wall}
{\sc \au{Marusic, I.}, \au{McKeon, B.~J.}, \au{Monkewitz, P.~A.}, \au{Nagib,
  H.~M.}, \au{Smits, A.~J.} \& \au{Sreenivasan, K.~R.}} \yr{2010}
  \at{{Wall-bounded turbulent flows at high Reynolds numbers: Recent advances
  and key issues}}.  \jt{Phys. Fluids}  \bvol{22}~(6),  \pg{065103}.

\bibitem[Moser {\em et~al.\/}(1999)Moser, Kim \& Mansour]{moser1999direct}
{\sc \au{Moser, R.~D.}, \au{Kim, J.} \& \au{Mansour, N.~N.}} \yr{1999}
  \at{{Direct numerical simulation of turbulent channel flow up to $Re_\tau=
  590$}}.  \jt{Phys. Fluids}  \bvol{11}~(4),  \pg{943--945}.

\bibitem[Niemela \& Sreenivasan(2008)]{niemela2008formation}
{\sc \au{Niemela, J.~J.} \& \au{Sreenivasan, K.~R.}} \yr{2008}  \at{{Formation
  of the “superconducting” core in turbulent thermal convection}}.
  \jt{Phys. Rev. Lett.}  \bvol{100}~(18),  \pg{184502}.

\bibitem[Nieuwstadt(1984)]{nieuwstadt1984turbulent}
{\sc \au{Nieuwstadt, F. T.~M.}} \yr{1984}  \at{{The turbulent structure of the
  stable, nocturnal boundary layer}}.  \jt{J. Atmos. Sci.}  \bvol{41}~(14),
  \pg{2202--2216}.

\bibitem[Pirozzoli {\em et~al.\/}(2017)Pirozzoli, Bernardini, Verzicco \&
  Orlandi]{pirozzoli2017mixed}
{\sc \au{Pirozzoli, S.}, \au{Bernardini, M.}, \au{Verzicco, R.} \& \au{Orlandi,
  P.}} \yr{2017}  \at{{Mixed convection in turbulent channels with unstable
  stratification}}.  \jt{J. Fluid Mech.}  \bvol{821},  \pg{482--516}.

\bibitem[Podvin \& Sergent(2015)]{podvin2015large}
{\sc \au{Podvin, B.} \& \au{Sergent, A.}} \yr{2015}  \at{{A large-scale
  investigation of wind reversal in a square Rayleigh--B{\'e}}nard cell}.
  \jt{J. Fluid Mech.}  \bvol{766},  \pg{172--201}.

\bibitem[Ricco {\em et~al.\/}(2012)Ricco, Ottonelli, Hasegawa \&
  Quadrio]{ricco2012changes}
{\sc \au{Ricco, P.}, \au{Ottonelli, C.}, \au{Hasegawa, Y.} \& \au{Quadrio, M.}}
  \yr{2012}  \at{{Changes in turbulent dissipation in a channel flow with
  oscillating walls}}.  \jt{J. Fluid Mech.}  \bvol{700},  \pg{77--104}.

\bibitem[Scagliarini {\em et~al.\/}(2015)Scagliarini, Einarsson, Gylfason \&
  Toschi]{scagliarini2015law}
{\sc \au{Scagliarini, A.}, \au{Einarsson, H.}, \au{Gylfason, {\'A}.} \&
  \au{Toschi, F.}} \yr{2015}  \at{{Law of the wall in an unstably stratified
  turbulent channel flow}}.  \jt{J. Fluid Mech.}  \bvol{781},  \pg{R5}.

\bibitem[Sharifnezhadazizi {\em et~al.\/}(2019)Sharifnezhadazizi, Norouzi,
  Prakash, Beale \& Khanbilvardi]{sharifnezhadazizi2019global}
{\sc \au{Sharifnezhadazizi, Z.}, \au{Norouzi, H.}, \au{Prakash, S.}, \au{Beale,
  C.} \& \au{Khanbilvardi, R.}} \yr{2019}  \at{{A global analysis of land
  surface temperature diurnal cycle using MODIS observations}}.  \jt{J. Appl.
  Meteorol. Climatol.}  \bvol{58}~(6),  \pg{1279--1291}.

\bibitem[Shishkina {\em et~al.\/}(2010)Shishkina, Stevens, Grossmann \&
  Lohse]{shishkina2010boundary}
{\sc \au{Shishkina, O.}, \au{Stevens, R.J.A.M.}, \au{Grossmann, S.} \&
  \au{Lohse, D.}} \yr{2010}  \at{{Boundary layer structure in turbulent thermal
  convection and its consequences for the required numerical resolution}}.
  \jt{New J. Phys.}  \bvol{12}~(7),  \pg{075022}.

\bibitem[Smits {\em et~al.\/}(2011)Smits, McKeon \& Marusic]{smits2011high}
{\sc \au{Smits, A.~J.}, \au{McKeon, B.~J.} \& \au{Marusic, I.}} \yr{2011}
  \at{{High--Reynolds number wall turbulence}}.  \jt{Annu. Rev. Fluid Mech.}
  \bvol{43}~(1),  \pg{353--375}.

\bibitem[Soucasse {\em et~al.\/}(2019)Soucasse, Podvin, Rivi{\`e}re \&
  Soufiani]{soucasse2019proper}
{\sc \au{Soucasse, L.}, \au{Podvin, B.}, \au{Rivi{\`e}re, P.} \& \au{Soufiani,
  A.}} \yr{2019}  \at{{Proper orthogonal decomposition analysis and modelling
  of large-scale flow reorientations in a cubic Rayleigh--B{\'e}}nard cell}.
  \jt{J. Fluid Mech.}  \bvol{881},  \pg{23--50}.

\bibitem[Stevens {\em et~al.\/}(2024)Stevens, Hartmann, Verzicco \&
  Lohse]{stevens2024wide}
{\sc \au{Stevens, R.J.A.M.}, \au{Hartmann, R.}, \au{Verzicco, R.} \& \au{Lohse,
  D.}} \yr{2024}  \at{{How wide must Rayleigh--B{\'e}nard cells be to prevent
  finite aspect ratio effects in turbulent flow?}}  \jt{J. Fluid Mech.}
  \bvol{1000},  \pg{A58}.

\bibitem[Stull(2015)]{stull2015practical}
{\sc \au{Stull, R.~B.}} \yr{2015} {\em {Practical meteorology: an algebra-based
  survey of atmospheric science}\/}.  \publ{University of British Columbia}.

\bibitem[Teimurazov {\em et~al.\/}(2023)Teimurazov, Singh, Su, Eckert,
  Shishkina \& Vogt]{teimurazov2023oscillatory}
{\sc \au{Teimurazov, A.}, \au{Singh, S.}, \au{Su, S.}, \au{Eckert, S.},
  \au{Shishkina, O.} \& \au{Vogt, T.}} \yr{2023}  \at{{Oscillatory large-scale
  circulation in liquid-metal thermal convection and its structural unit}}.
  \jt{J. Fluid Mech.}  \bvol{977},  \pg{A16}.

\bibitem[Urban {\em et~al.\/}(2022)Urban, Hanzelka, Kr{\'a}lik, Musilov{\'a} \&
  Skrbek]{urban2022thermal}
{\sc \au{Urban, P.}, \au{Hanzelka, P.}, \au{Kr{\'a}lik, T.}, \au{Musilov{\'a},
  V.} \& \au{Skrbek, L.}} \yr{2022}  \at{{Thermal waves and heat transfer
  efficiency enhancement in harmonically modulated turbulent thermal
  convection}}.  \jt{Phys. Rev. Lett.}  \bvol{128}~(13),  \pg{134502}.

\bibitem[Vogt {\em et~al.\/}(2018)Vogt, Horn, Grannan \& Aurnou]{vogt2018jump}
{\sc \au{Vogt, T.}, \au{Horn, S.}, \au{Grannan, A.~M.} \& \au{Aurnou, J.~M.}}
  \yr{2018}  \at{{Jump rope vortex in liquid metal convection}}.  \jt{Proc.
  Natl. Acad. Sci. U. S. A.}  \bvol{115}~(50),  \pg{12674--12679}.

\bibitem[Wang {\em et~al.\/}(2020)Wang, Zhou \& Sun]{wang2020vibration}
{\sc \au{Wang, B.-F.}, \au{Zhou, Q.} \& \au{Sun, C.}} \yr{2020}
  \at{{Vibration-induced boundary-layer destabilization achieves massive
  heat-transport enhancement}}.  \jt{Sci. Adv.}  \bvol{6}~(21),  \pg{eaaz8239}.

\bibitem[Wu {\em et~al.\/}(2021)Wu, Dong, Wang \& Zhou]{wu2021phase}
{\sc \au{Wu, J.-Z.}, \au{Dong, Y.-H.}, \au{Wang, B.-F.} \& \au{Zhou, Q.}}
  \yr{2021}  \at{{Phase decomposition analysis on oscillatory
  Rayleigh--B{\'e}}nard turbulence}.  \jt{Phys. Fluids}  \bvol{33}~(4),
  \pg{045108}.

\bibitem[Wu {\em et~al.\/}(2022)Wu, Wang \& Zhou]{wu2022massive}
{\sc \au{Wu, J.-Z.}, \au{Wang, B.-F.} \& \au{Zhou, Q.}} \yr{2022}  \at{{Massive
  heat transfer enhancement of Rayleigh-B{\'e}nard turbulence over rough
  surfaces and under horizontal vibration}}.  \jt{Acta Mech. Sin.}
  \bvol{38}~(2),  \pg{321319}.

\bibitem[Xia {\em et~al.\/}(2023)Xia, Huang, Xie \& Zhang]{xia2023tuning}
{\sc \au{Xia, K.-Q.}, \au{Huang, S.-D.}, \au{Xie, Y.-C.} \& \au{Zhang, L.}}
  \yr{2023}  \at{{Tuning heat transport via coherent structure manipulation:
  recent advances in thermal turbulence}}.  \jt{Natl. Sci. Rev.}
  \bvol{10}~(6),  \pg{nwad012}.

\bibitem[Xu {\em et~al.\/}(2021)Xu, Chen \& Xi]{xu2021tristable}
{\sc \au{Xu, A.}, \au{Chen, X.} \& \au{Xi, H.-D.}} \yr{2021}  \at{{Tristable
  flow states and reversal of the large-scale circulation in two-dimensional
  circular convection cells}}.  \jt{J. Fluid Mech.}  \bvol{910},  \pg{A33}.

\bibitem[Xu \& Li(2023)]{xu2023multi}
{\sc \au{Xu, A.} \& \au{Li, B.-T.}} \yr{2023}  \at{{Multi-GPU thermal lattice
  Boltzmann simulations using OpenACC and MPI}}.  \jt{Int. J. Heat Mass
  Transf.}  \bvol{201},  \pg{123649}.

\bibitem[Xu {\em et~al.\/}(2019)Xu, Shi \& Xi]{xu2019lattice}
{\sc \au{Xu, A.}, \au{Shi, L.} \& \au{Xi, H.-D.}} \yr{2019}  \at{{Lattice
  Boltzmann simulations of three-dimensional thermal convective flows at high
  Rayleigh number}}.  \jt{Int. J. Heat Mass Transf.}  \bvol{140},
  \pg{359--370}.

\bibitem[Xu {\em et~al.\/}(2017)Xu, Shi \& Zhao]{xu2017accelerated}
{\sc \au{Xu, A.}, \au{Shi, L.} \& \au{Zhao, T.~S.}} \yr{2017}  \at{{Accelerated
  lattice Boltzmann simulation using GPU and OpenACC with data management}}.
  \jt{Int. J. Heat Mass Transf.}  \bvol{109},  \pg{577--588}.

\bibitem[Yang {\em et~al.\/}(2020)Yang, Chong, Wang, Verzicco, Shishkina \&
  Lohse]{yang2020periodically}
{\sc \au{Yang, R.}, \au{Chong, K.~L.}, \au{Wang, Q.}, \au{Verzicco, R.},
  \au{Shishkina, O.} \& \au{Lohse, D.}} \yr{2020}  \at{{Periodically modulated
  thermal convection}}.  \jt{Phys. Rev. Lett.}  \bvol{125}~(15),  \pg{154502}.

\bibitem[Yao {\em et~al.\/}(2022)Yao, Chen \& Hussain]{yao2022direct}
{\sc \au{Yao, J.}, \au{Chen, X.} \& \au{Hussain, F.}} \yr{2022}  \at{{Direct
  numerical simulation of turbulent open channel flows at moderately high
  Reynolds numbers}}.  \jt{J. Fluid Mech.}  \bvol{953},  \pg{A19}.

\bibitem[Yerragolam {\em et~al.\/}(2024)Yerragolam, Howland, Stevens, Verzicco,
  Shishkina \& Lohse]{yerragolam2024scaling}
{\sc \au{Yerragolam, G.S.}, \au{Howland, C.J.}, \au{Stevens, R.J.A.M.},
  \au{Verzicco, R.}, \au{Shishkina, O.} \& \au{Lohse, D.}} \yr{2024}
  \at{{Scaling relations for heat and momentum transport in sheared
  Rayleigh-B{\'e}nard convection}}.  \jt{J. Fluid Mech.}  \bvol{1000},
  \pg{A74}.

\bibitem[Yerragolam {\em et~al.\/}(2022)Yerragolam, Verzicco, Lohse \&
  Stevens]{yerragolam2022small}
{\sc \au{Yerragolam, G.S.}, \au{Verzicco, R.}, \au{Lohse, D.} \& \au{Stevens,
  R.J.A.M.}} \yr{2022}  \at{How small-scale flow structures affect the heat
  transport in sheared thermal convection}.  \jt{J. Fluid Mech.}  \bvol{944},
  \pg{A1}.

\bibitem[Young {\em et~al.\/}(2002)Young, Kristovich, Hjelmfelt \&
  Foster]{young2002rolls}
{\sc \au{Young, G.~S.}, \au{Kristovich, D. A.~R.}, \au{Hjelmfelt, M.~R.} \&
  \au{Foster, R.~C.}} \yr{2002}  \at{{Rolls, streets, waves, and more: A review
  of quasi-two-dimensional structures in the atmospheric boundary layer}}.
  \jt{Bull. Amer. Meteorol. Soc.}  \bvol{83}~(7),  \pg{997--1002}.

\bibitem[Zhang(2024)]{zhang2024structure}
{\sc \au{Zhang, H.}} \yr{2024}  \at{{Structure and coupling characteristics of
  multiple fields in dust storms}}.  \jt{Acta Mech. Sin.}  \bvol{40},
  \pg{123339}.

\bibitem[Zhang {\em et~al.\/}(2023)Zhang, Tan \& Zheng]{zhang2023multifield}
{\sc \au{Zhang, H.}, \au{Tan, X.-L.} \& \au{Zheng, X.-J.}} \yr{2023}
  \at{Multifield intermittency of dust storm turbulence in the atmospheric
  surface layer}.  \jt{J. Fluid Mech.}  \bvol{963},  \pg{A15}.

\bibitem[Zonta {\em et~al.\/}(2022)Zonta, Sichani \&
  Soldati]{zonta2022interaction}
{\sc \au{Zonta, F.}, \au{Sichani, P.~H.} \& \au{Soldati, A.}} \yr{2022}
  \at{{Interaction between thermal stratification and turbulence in channel
  flow}}.  \jt{J. Fluid Mech.}  \bvol{945},  \pg{A3}.

\end{thebibliography}

\end{document}